\documentclass[aps,prb,floatfix,superscriptaddress]{revtex4-2}
\usepackage{graphicx,amsfonts}
\usepackage{bm,color}
\usepackage{multirow}
\usepackage{amssymb,amsmath,hyperref}
\usepackage{wrapfig}
\usepackage{float}
\DeclareUnicodeCharacter{2212}{-}

\usepackage{amsmath,amsfonts,amssymb}
\usepackage{hyperref}
\hypersetup{colorlinks=true, linkcolor=blue, citecolor=red, urlcolor=blue}
\usepackage[utf8]{inputenc}
\usepackage{mathtools}
\usepackage{braket}
\usepackage{mathtools}
\usepackage{braket}
\usepackage{circuitikz}
\usepackage{nicefrac,xfrac}
\usepackage[table]{xcolor}
\RequirePackage[normalem]{ulem} 
 
\newcommand{\Tr}{{\rm Tr}}
\newcommand{\TrB}{{\rm Tr}_{\rm B}}
\newcommand{\id}{{\mathbf{1}}}
\newcommand{\comment}[1]{}

\newcommand{\ii}{{\rm i}}
\newcommand{\be}{\begin{equation}\begin{aligned}\label}{}
\newcommand{\ee}{\end{aligned}\end{equation}}

\begin{document}

\title{Heat transport in driven quantum systems: Generalized Floquet-Redfield equation and  master equation in the instantaneous eigenbasis} 

\author{Luca Magazz\`u}
\affiliation{Pico group, Department of Applied Physics, Aalto University School of Science, P.O. Box 13500, 00076 Aalto, Finland}

\author{Christoforus Dimas Satrya}
\affiliation{Pico group, Department of Applied Physics, Aalto University School of Science, P.O. Box 13500, 00076 Aalto, Finland}
\affiliation{Department of Microtechnology and Nanoscience, Chalmers University of Technology, 412 96 Gothenburg, Sweden}

\author{Aleksandr S. Strelnikov}
\affiliation{Pico group, Department of Applied Physics, Aalto University School of Science, P.O. Box 13500, 00076 Aalto, Finland}

\author{Bayan Karimi}
\affiliation{Pico group, Department of Applied Physics, Aalto University School of Science, P.O. Box 13500, 00076 Aalto, Finland}
\affiliation{Pritzker School of Molecular Engineering, University of Chicago, Chicago IL 60637, USA}

\author{Jukka P. Pekola}
\affiliation{Pico group, Department of Applied Physics, Aalto University School of Science, P.O. Box 13500, 00076 Aalto, Finland}

\date{\today}

\begin{abstract}
We provide a comprehensive study of heat transport in periodically driven quantum systems using a master equation approach in the Floquet basis. Starting from the exact, formal expressions, we obtain, within the weak coupling regime of system-bath coupling, a generalized Floquet-Redfield master equation which does not involve Markovian or secular approximations and the corresponding formula for the heat current. From this approach we also derive the  Markovian Floquet-Redfield equation and its full-secular approximation. Moreover, we present a master equation for the driven spin-boson model in the instantaneous eigenbasis. 
The different approaches are compared by applying them to the steady-state heat transport in the driven spin-boson model and the impact of the different approximations is assessed. An analytical solution is provided for the master equation in the instantaneous eigenbasis which reproduces the numerics and explains the features of the heat current as a function of the drive parameters.  
\end{abstract}

\maketitle

\section{Introduction}

The theory of periodically driven open quantum systems~\cite{Davies1978, Kohler1997,Grifoni1998,Fonseca-Romero2004,Hone2009,Deng2016,Magazzu2018PRE,Reimer2018,Menczel2019,Elouard2020,Teixeira2022,Mori2023,DiMeglio2024,Cao2026,Akbari2026} provides the tools for modeling the behavior and engineering the response of quantum devices subject to dissipation and driven by external, time-periodic controls. In particular, for quantum thermal devices, the environment is an essential part of the experimental platforms, and can be embodied by engineered heat baths with controlled temperatures and couplings to the system.  
The rapid  advancement of these devices~\cite{Pekola2015,Campbell2026,Satrya2026} poses the problem of controlling the heat flow between the quantum system and the environment via time-dependent external fields, especially in view of realizing quantum-thermodynamical machines~\cite{Karimi2016,Restrepo2018,Pekola2019,Funo2019,Uusnakki2026} characterized by periodic change of the Hamiltonian parameters. 
In this situation, the system's response is determined by the interplay between dissipation and external driving. For example, the latter can be tuned to suppress the system dynamics, via the mechanism of coherent destruction of tunneling~\cite{Grossmann1991} or to effectively decouple the system from the environment~\cite{Viola1998}. In turn, this response  governs transport through the quantum system~\cite{Kohler2005,Rehman2026}. While most of the studies on driven open systems have focused on spectroscopy~\cite{Oliver2005,Sillanpaeae2006,Koski2018,Chen2021, Magazzu2021,Gu2023}, here we consider the problem of heat transport using the master equation approach. This approach is best suited for situations where the system and the heat baths exchange energy in the regime of weak coupling, with no limitations on the degree of anharmonicity of the system.\\ 
\indent A system subject to an external, classical driving field of frequency $\omega_{\rm d}$
is usually treated according to a Markovian, weak coupling master equation. 
The standard form is the Gorini-Kossakowski-Sudarshan-Lindblad master equation~\cite{Breuer2002} in the Floquet basis, namely the Floquet-Lindblad master equation~\cite{Alicki2012,Alicki2013,
Ikeda2020,Clawson2025,Ilinskaya2026, Keliri2026}. This equation is derived starting from the Born-Markov master equation for the driven system and performing the secular approximation in the basis of Floquet states, characterized by the quasienergies $\{\varepsilon_n\}$, with respect to the \emph{extended} Bohr frequencies $\varepsilon_n-\varepsilon_m + k\omega_{\rm d}$.  
Since approximations to the Born master equation can have a different regimes of validity in the presence of time-dependent driving with respect to the static case~\cite{DiMeglio2024}, and can impact the thermodynamic consistency of the results~\cite{Tude2026}, it is valuable to have a precise and manageable master equation which treats the driving exactly and does not resort to Markovian or secular approximations. 

In this work, we start from the exact Nakajima-Zwanzig equation for the time evolution of the reduced density matrix (RDM) of an open system described by a time-dependent Hamiltonian. For a time-periodic driving, assuming the periodicity of the steady state and expanding to leading order in the system-bath coupling (born approximation) the dissipator of this exact, time-nonlocal equation, and finally projecting in the basis of Floquet states, we obtain a time-local, weak-coupling master equation for the Fourier components of the RDM, the steady-state generalized Floquet-Redfield (GFR) equation. An analogous procedure yields a weak-coupling, but otherwise exact, expression for the Fourier components of the time-periodic steady-state heat current to a thermal bath, which requires as input the solution for the RDM. This approach, treats the drive exactly and does not rely on Markovian or secular approximations. We also derive the Markovian-approximated version of the GFR equation, the Floquet-Redfield equation, and the corresponding expression of the heat current.
We note that nonsecular master equations do not generally guarantee complete positivity. Nevertheless, in~\cite{Hartmann2020} it is argued that the Redfield-type master equations give unphysical results when applied beyond their regime of validity: further approximations enforcing complete positivity (as in Lindblad-type master equations) do not necessarily improve accuracy. Within their regimes of validity, Redfield-type master equations can provide more accurate results than secularized master equations enforcing complete positivity. In the case of driven system, the quasienergy spacing may become very small: this is incompatible with the secular approximation which potentially leads to strong artifacts~\cite{Hotz2021}.
The full secular approximation, applied to the GFR, yields a standard Pauli-type equation for the populations of the system. These master equations are compared with an application to the driven spin-boson model~\cite{Weiss2012}, namely a driven qubit coupled to a bosonic heat bath, highlighting the role of the steady-state coherences in the basis of Floquet states. A comparison is also made between the GFR equation and its Markovian-approximated version: As expected, their results for the heat current coincide in the ultra-weak system-bath coupling limit. However, their predictions for the heat current disagree at finite coupling, with those of the non-Markovian GFR being confirmed by HEOM evaluations.\\
\indent The Markov approximation also yields a master equation for the transient dynamics with a time-dependent dissipator. Starting from this, a master equation in the instantaneous eigenbasis~\cite{Pekola2016,Karimi2016,KarimiPhDthesis} is also presented and applied to the spin-boson model upon propagating the solution up to the time-periodic steady state. The resulting  steady-state heat current is compared to that of the GFR approach, used as a benchmark, at the ultra-weak coupling. The results of the two approaches show agreement, especially at low drive frequency, with deviations at high frequency, away from the validity regime of the derivation of the master equation in the instantaneous eigenbasis.\\
\indent We finally solve analytically the master equation in the instantaneous eigenbasis by resorting to  further approximations. The analytical solution essentially reproduces the results from the numerics in the whole frequency range, for not too large drive amplitude, and allows us to explain in simple terms the characteristics of the heat current peaks found when the drive frequency matches integer fractions of the average qubit frequency. These features stem from the physics of the driven spin-boson model, which involves multi-photon processes yielding peaks in the heat current at drive frequencies that match fractions of the qubit frequency (renormalized by the drive): In the absence of applied static bias, the peaks at even fractions are suppressed~\cite{Grifoni1995,Ho1986, Shytov2003,Shevchenko2005,Deppe2008,Thomas2023} according to a symmetry-induced selection rule. The methods considered in this work capture all these features of the driven spin-boson model.\\

\section{Generalized Floquet-Redfield (GFR) approach to heat transport in a driven quantum system}
\label{Floquet-Redfield}

\subsection{Nakajima-Zwanzig equation approach}

Consider the model of a driven system coupled to a number of independent heat baths according to the total Hamiltonian
\be{H_total}
\hat H(t)=\tilde H_{\rm S}(t)+ \hat H_{\rm B} + \hat H_{\rm SB}\;,
\ee
as detailed in Appendix~\ref{CaldeiraLeggett_model}. Here, $\tilde H_{\rm S}(t)$ is a generic, time-dependent system Hamiltonian, including the bath-induced counterterm prescribed by the Caldeira-Leggett Hamiltonian, and $\hat H_{\rm B}=\sum_{{\rm b},j}\hat H_{\rm b}=\sum_{{\rm b},j}\hbar \omega_{\rm b}(\hat b^\dag_{{\rm b}j}\hat b_{{\rm b}j}+1/2)$ where the index ${\rm b}$ runs over the baths.  Assume that the system couples to bath ${\rm b}$ via the system operator $\hat Q_{\rm b}$. Then we set  $\hat H_{\rm SB}=\sum_{\rm b}\hat Q_{\rm b}\otimes\hat B_{\rm b}$, where $\hat B_{\rm b}$ is a collective bath operator, e.g, for position coupling $\hat B_{\rm b}=\sum_j\hbar\lambda_{{\rm b}j}(\hat b_{{\rm b}j}+\hat b_{{\rm b}j}^\dag)$, while for momentum coupling $\hat{B}_{\rm b}:=\ii\hbar\sum_j\lambda_{{\rm b}j}(\hat b_{{\rm b}j}^\dag - \hat b_{{\rm b}j})$ .\\
\indent We assume a factorized initial condition for the total density matrix $\varrho(t)$, namely $\varrho(0)=\rho(0)\otimes \rho_{\rm B}$, with $\rho(t)$ the system reduced density matrix (RDM) at time $t$ and $\rho_{\rm B}=\bigotimes_{\rm b}\exp(-\beta_{\rm b}\hat H_{\rm b})/Z_{\rm b}$ the tensor product of thermal states of the baths at inverse temperatures $\beta_{\rm b}$ with partition functions $Z_{\rm b}={\rm Tr}\exp(-\beta_{\rm b}\hat H_{\rm b})$. The generalized master equation for the driven dissipative system can be obtained in a formally exact fashion by using the projection operator technique and has the form of the Nakajima-Zwanzig equation~\cite{Nakajima1958,Zwanzig1960,Meier1999,Donarini2024,Magazzu2024PRB,Gonzalez-Ballestero2024} (see Appendix~\ref{NZ_derivation} for a detailed derivation)
\be{NZ_equation}
\dot\rho(t) = \mathcal{L}_{\rm S}(t)\rho(t) + \int_0^t dt'  \mathcal{K}(t,t')\rho(t')\;,
\ee
where the kernel superoperator acts on the RDM according to $\mathcal{K}(t,t')\rho(t')={\rm Tr}_{\rm B}\{\mathcal{L}_{\rm SB} \mathcal{G}_{\mathcal{Q}}(t,t')\mathcal{L}_{\rm SB}\rho(t')\otimes\rho_{\rm B}\}$. The Liouville superoperators are defined as $\mathcal{L}_i \bullet=-(\ii/\hbar)[\hat H_i,\bullet]$, with $\hat H_i$ given in Eq.~\eqref{H_total}. \\
\indent To leading (zeroth) order in the system-bath coupling, the irreducible propagator has the explicit expression $\mathcal{G}_{\mathcal{Q}}^{(0)}(t,t')\bullet=U_{\rm S}(t,t')U_{\rm B}(t-t')\bullet U_{\rm S}^\dag(t,t')U_{\rm B}^\dag(t-t')$~\cite{Magazzu2024PRB,Donarini2024}. This expression gives the leading-order kernel superoperator
\be{NZ_kernel_second}
\mathcal{K}^{(2)}(t,t')\rho(t')=&
-\frac{1}{\hbar^2}\sum_{\rm b}\Big\{\left[\hat{Q}_{\rm b}U_{\rm S}(t,t')\hat{Q}_{\rm b}\rho(t')U_{\rm S}^\dag(t,t')-U_{\rm S}(t,t')\hat{Q}_{\rm b}\rho(t')U_{\rm S}^\dag(t,t')\hat{Q}_{\rm b}\right]C_{\rm b}(t-t')\\
&\;\qquad\qquad+\left[U_{\rm S}(t,t')\rho(t')\hat{Q}_{\rm b}U_{\rm S}^\dag(t,t')\hat{Q}_{\rm b}-\hat{Q}_{\rm b}U_{\rm S}(t,t')\rho(t')\hat{Q}_{\rm b}U_{\rm S}^\dag(t,t')\right]C_{\rm b}^*(t-t')\Big\}\;,
\ee
where, introducing the bath spectral density function $G_{\rm b}(\omega)=\sum_j\lambda_{{\rm b}j}^2\delta(\omega-\omega_{_{\rm b}j})$, the bath correlation function of bath ${\rm b }$ reads 
\be{C_t}
C_{\rm b}(t):=&\;\langle \hat B_{\rm b}(t)\hat B_{\rm b}(0)\rangle\\
=&\;\hbar^2\int_0^\infty d \omega\;G_{\rm b}(\omega)\left[\coth\left(\frac{\beta_{\rm b}\hbar\omega}{2}\right)\cos(\omega t)-{\rm i}\sin(\omega t)\right]\;,
\ee
see Appendix~\ref{CaldeiraLeggett_model}.
The explicit expression for the system evolution operator is
\be{}
U_{\rm S}(t,t')=\mathcal{T}e^{-\frac{\rm i}{\hbar}\int_{t'}^t d\tau\;\hat H_{\rm S}(\tau)}\;.
\ee
Note that, in view of a second-order treatment of the system-bath interaction, in this definition we use the bare system Hamiltonian $\hat H_{\rm S}(t)$, without the counterterm of second order in the system-bath coupling, see Eq.~\eqref{A2}, prescribed by the Caldeira-Leggett model. Consistently, the unitary evolution term in the Born approximation of Eq.~\eqref{NZ_equation} retains this term.

The Born equation, namely Eq.~\eqref{NZ_equation}, endowed with the second-order kernel in Eq.~\eqref{NZ_kernel_second}, is time-nonlocal. The only approximation enforced is leading (second) order in the system-bath coupling. The usual  Markovian approximation, where, in the interaction picture, one sets $\rho_{\rm I}(t-\tau)\simeq\rho_{\rm I}(t)$, see Appendix~\ref{Markovian}, is not necessary for the steady state, provided that the memory kernel is integrable. As shown below, even for a periodically driven system, the steady-state  master equation in Fourier space is time-local, with no need to enforce Markovianity. Alternative perturbative approaches which treat the drive exactly and are suited for the strong system-bath coupling regime are given, e.g.  in~\cite{Grifoni1995,Magazzu2018,Magazzu2021}.

\subsubsection{Heat current}
\label{heat_current}

The current operator yielding the heat current to bath ${\rm b}$ is defined as $\hat I_{\text b}=\frac{\ii}{\hbar}[\hat H(t), \hat H_{\rm b}]=\frac{\ii}{\hbar}[\hat H_{\rm SB}, \hat H_{\rm b}]=\hat Q_{\rm b}\otimes \bar{B}_{\rm b}$. Its expectation value $P_{\rm b}(t)={\rm Tr}[\hat I_{\rm b}\varrho(t)]=\frac{d}{dt}{\rm Tr}[\hat H_{\rm b}\varrho(t)]$ equals the rate of change of the bath energy. The current operator has a similar structure as the system-bath coupling term, reading $\hat H_{\rm SB}=\sum_{\rm b}\hat Q_{\rm b}\otimes \hat{B}_{\rm b}$. Bath-position coupling, see Eq.~\eqref{Hterms}, yields the bath current operator $\bar{B}_{\rm b}:=\frac{\ii}{\hbar}[\hat B_{\rm b},\hat H_{\rm b}]={\rm i} \hbar\sum_{j}\lambda_{{\rm b}j}\omega_{{\rm b}j}(\hat b_{{\rm b}j} - \hat b^\dag_{{\rm b}j})$. For bath-momentum coupling, see Eq.~\eqref{HSB_momentum}, we have $\bar{B}_{\rm b}:=\hbar\sum_{j}\lambda_{{\rm b}j}\omega_{{\rm b}j}(\hat b_{{\rm b}j} + \hat b^\dag_{{\rm b}j})$.
In either case the heat current to the bath $\rm b$ has the exact expression~\cite{Magazzu2024PRB,Donarini2024}
\be{Pb_exact}
P_{\rm b}(t)&={\rm Tr}_{\rm S}\int_0^t dt' \; \mathcal{K}_{I,{\rm b}}(t,t')\rho(t')\;,
\ee
where $\mathcal{K}_{I,{\rm b}}(t,t')\rho(t')={\rm Tr}_{\rm B}\{\hat I_{\text b} \mathcal{G}_{\mathcal{Q}}(t,t')\mathcal{L}_{\rm SB}\rho(t')\otimes\rho_B\}$. Details are provided in Appendix~\ref{NZ_derivation}.\\
\indent To leading (second) order in the system-bath coupling, we take $\mathcal{G}_{\mathcal{Q}}(t,t')\simeq \mathcal{G}_{\mathcal{Q}}^{(0)}(t,t')$, as for the kernel of the generalized master equation, see Eq.~\eqref{NZ_kernel_second}, and the resulting explicit expression for the action of the current kernel is
\be{KI_second}
\mathcal{K}^{(2)}_{I,{\rm b}}(t,t')\rho(t')=& -\frac{\ii}{\hbar}\Big\{\hat{Q}_{\rm b} U_{\rm S}(t,t')\hat{Q}_{\rm b} \rho(t') U_{\rm S}^\dag(t,t') K_{\rm b}(t-t')\\
&\qquad 
-\hat{Q}_{\rm b} U_{\rm S}(t,t')\rho(t')\hat{Q}_{\rm b}  U_{\rm S}^\dag(t,t') K_{\rm b}^*(t-t')\Big\}\;.
\ee
Here, we introduced the correlator~\footnote{Note that, as $\bar{B}_{\rm b}$ is Hermitian, $\langle \hat B_{\rm b}(t')\bar{B}_{\rm b}(t)\rangle=K_{\rm b}^*(t-t')$.} 
\be{}
K_{\rm b}(t)=&\;\langle \bar{B}_{\rm b}(t)\hat B_{\rm b}(0)\rangle\\
=&\;\hbar^2\int_0^\infty d \omega\;\omega G_{\rm b}(\omega)\left[\coth\left(\frac{\beta_{\text b}\hbar\omega}{2}\right)\sin(\omega t)+{\rm i}\cos(\omega t)\right]
\equiv \;-\frac{d}{dt}C_{\rm b}(t) \;,
\ee
with $C_{\rm b}(t)$ the bath correlation function of bath ${\rm b}$, see Eq.~\eqref{C_t}.\\
\indent We conclude this section by noting that the definition of heat current to the bath given here reflects a specific partition of the super-system into system, bath, and interaction terms. Other choices are possible with different results in the transient, especially at strong coupling, where the energy stored in the system-bath interaction is non-negligible~\cite{Carrega2016}. At the periodic steady state, the definition given here yields, $\sum_{\rm b}\overline{ P_{\rm b}(t)}=\overline{\dot {W}(t)}$, where $W$ is the work done on the system by the drive, namely $\dot W(t)={\rm Tr}_{\rm S}\{\rho(t)\partial_t\hat{H}_{\rm S}(t)\}$, and where $\overline{\bullet}$ denotes one-period averaging. Indeed, using the Liouville-von Neumann equation for the evolution of the total system one can show that $\frac{d}{dt}\langle\hat H(t)\rangle=\langle\partial_t\hat H_{\rm S}(t)\rangle=\dot W(t)$. On the other hand $\frac{d}{dt}\langle\hat H(t)\rangle=\frac{d}{dt}\langle\hat H_{\rm B}\rangle + \frac{d}{dt}[E_{\rm S}(t) + E_{\rm SB}(t)]$ where $\frac{d}{dt}\langle\hat H_{\rm B}\rangle=\sum_{\rm b}P_{\rm b}(t)$ and where $E_i(t)=\langle \hat H_i(t)\rangle$. Since at the periodic steady state the one-period average of the stored system and interaction energies vanishes, we arrive at the equality $\sum_{\rm b}P_{\rm b}^{(0)}=\overline{\dot W}$, where $P_{\rm b}^{(0)}$ is the zero-frequency coefficient of the Fourier expansion of the periodic asymptotic heat current to bath ${\rm b}$.

\subsection{Floquet theory}

For a time-periodic system Hamiltonian with period $T=2\pi/\omega_{\text d}$, it is convenient to project the generalized master equation~\eqref{NZ_equation}-\eqref{NZ_kernel_second} in the basis of Floquet states $|\psi_n(t)\rangle$. These are the solutions of the Schr\"odinger equation for closed systems with time-periodic Hamiltonian $\tilde{H}_{\rm S}(t+T)=\tilde{H}_{\rm S}(t)$, namely 
\be{SEFloquet}
\ii\hbar\partial_t|\Psi_n(t)\rangle = \tilde{H}_{\rm S}(t)|\Psi_n(t)\rangle\;.
\ee
The Floquet theorem~\cite{Floquet1879,Grifoni1998} states that these solutions are given by
\be{Floquet_st}
|\Psi_n(t)\rangle=e^{-\ii \varepsilon_n t}|u_n(t)\rangle\;.
\ee
The time-periodic \emph{Floquet modes} $|u_n(t)\rangle=|u_n(t+T)\rangle$, form a time-dependent basis of the Hilbert space of the system. Along with the quasienergies $\varepsilon_n$, they solve the eigenvalue problem
\be{QE}
[\tilde{H}_{\rm S}(t)-{\rm i}\hbar\partial_t]|u_n(t)\rangle=\hbar\varepsilon_n |u_n(t)\rangle\;.
\ee
Note that the Floquet state with Floquet mode $e^{-\ii m\omega_{\rm d}t}\ket{u_n(t)}$ is equivalent to the state with mode $\ket{u_n(t)}$ and quasienergy $\varepsilon_n + m\omega_{\rm d}$. Therefore, the quasienergies are organized in bands with replicas separated by the drive frequency $\omega_{\rm d}$. 
A note on an efficient numerical implementation of the Floquet eigensystem is provided in Appendix~\ref{numerical_implementation}.
In the static limit, $\varepsilon_n \rightarrow E_n/\hbar$ and $\ket{u_n(t)}\rightarrow \ket{n}$ so that Eq.~\eqref{QE} becomes $\tilde H_{\rm S}\ket{n}=\hbar\omega_n\ket{n}$. Throughout this work we use the definition 
\be{omega_nm}
\omega_{nm}:=\varepsilon_n-\varepsilon_m\;.
\ee
\indent To obtain the master equation in the basis of Floquet states, we use the fact that the latter solve the Schr\"odinger equation, namely 
\be{floquet_propagator}
U_{\rm S}(t_2,t_1)|\Psi_n(t_1)\rangle=|\Psi_n(t_2)\rangle\;.
\ee
Omitting the bath index, the matrix elements of the system coupling operator in the Floquet states basis can be written as $Q_{n m }(t)=\bra{\Psi_n(t)}\hat Q\ket{\Psi_m(t)}=\tilde Q_{nm}(t)e^{{\rm i }\omega_{n m  }t}$, where $\tilde Q_{nm}  (t)=\bra{u_n (t)}\hat{Q}\ket{u_m  (t)}$.
Expanding the Floquet modes in Fourier series
$|u_{n }(t)\rangle=\sum_k e^{{\rm i}k\omega_{\rm d} t} |u_n ^{(k)}\rangle$
yields for the matrix element of the system operator $\hat{Q}$ in the basis of Floquet modes
\be{result1}
\tilde Q_{nm}(t)
&=\sum_{k,k'}e^{\ii(k'-k) {\omega_{\text d}} t}\bra{u_n ^{(k)}}\hat{Q}\ket{u_m  ^{(k')}}\\
&=\sum_{l}e^{\ii l {\omega_{\text d}} t}Q_{n m  }^{l}\;,
\ee
where $Q_{n m  }^{l}:=\sum_{k}\bra{u_n ^{(k)}}\hat{Q}\ket{u_m  ^{(k+l)}}$, satisfying the symmetry $Q_{n   m  }^{-l}={(Q_{m  n  }^{l})^*}$.

\subsection{Exact results at the steady state}

Contrary to the static case, where a heat current at the steady state requires multiple baths with a temperature bias, in the presence of a time-dependent driving, a steady-state heat current  can be induced by the drive on the system, even in absence of additional baths or of temperature biases.\\
\indent Throughout this work, ``steady state'' denotes the asymptotic periodic dynamics 
with the period $\tau_{\rm d}=2\pi/\omega_{\rm d}$ of the drive. We assume convergence to a unique periodic solution whose existence and
uniqueness are not established here in general.
Moreover, here and in what follows, we consider the case in which the driven open system is coupled to a single bath. As a result, the bath index $\rm b$ and the sum over the baths are removed. The general case of multiple, independent baths can be recovered by reintroducing the bath index in the system operators $\hat Q$ and bath correlation function $C(t)$, and summing over $\rm b$ where appropriate, see e. g.  Appendix~\ref{Markovian}.\\
\indent We start form the exact Nakajima-Zwanzig equation~\eqref{NZ_equation}. To investigate the steady state-solution for the RDM for a time-periodic driving on the system, it is convenient to project Eq.~\eqref{NZ_equation} in the basis of Floquet modes. 
To do so, we insert the resolution of the identity $\mathbf{1}=\sum_n |\Psi_n(t)\rangle\langle \Psi_n(t)|$, with quasienergies within a given Brillouin zone, e.g. $\varepsilon_n\in [-\omega_{\rm d}/2,\omega_{\rm d}/2]$, and the decomposition of the RDM in the Floquet basis 
$\rho(t)=\sum_{n,m}\rho_{nm}(t)\ket{\Psi_n(t)}\bra{\Psi_m(t)}$.  
Then the matrix elements of the RDM read
$\rho_{nm}(t)=e^{i\omega_{nm}(t)}\tilde \rho_{nm}(t)$, where
$\tilde\rho_{nm}(t)=\bra{u_n(t)}\rho(t)\ket{u_m(t)}$.
Assume that the elements of the kernel superoperator vanish for large differences $\tau =t-t'$, and that the elements of the steady-state RDM are time-periodic (possibly constant): $\tilde\rho_{nm}(t\rightarrow\infty)=\sum_k e^{\ii k\omega_{\rm d} t}\tilde\rho^{(k)}_{nm}$. Defining the time-dependent tensor 
$$ R^{k}_{nmn'm'}(t) := \lim_{t\to \infty} \int_0^t d\tau \bra{u_n(t)}\mathcal{K}(t,t-\tau)[\ket{u_{n'}(t-\tau)}\bra{u_{m'}(t-\tau)} ]\ket{u_{m}(t)}e^{\ii k\omega_{\rm d}(t-\tau)}$$
and using Eq.~\eqref{QE}, we find the steady-state, time-local master equation
\be{NZ_Fourier}
\partial_t\tilde\rho_{nm}(t)=&  
-\ii\omega_{nm}\tilde\rho_{nm}(t)+\sum_{n'm',k'}R_{nmn'm'}^{k'}(t)\tilde\rho^{(k')}_{n'm'}\;.
\ee
This result is exact and describes the steady state of a periodically driven open quantum system. 

\subsection{Steady-state heat transport at weak system-bath coupling}
\label{sec_GFR}

\indent In the weak system-bath coupling limit, using the second-order expression~\eqref{NZ_kernel_second} for $\mathcal{K}(t,t')$ and Eq.~\eqref{floquet_propagator} yields the steady-state dissipation tensor
\be{FR_tensor} 
R^{k'(2)}_{nmn'm'}(t)=& -\frac{1}{\hbar^2}\int_0^\infty d\tau\Bigg\{\sum_p\Big[\delta_{m  m'}\tilde Q_{n p }(t)\tilde Q_{p  n'}(t-\tau)e^{-{\rm i }\omega_{pm'}\tau}C(\tau)+ \delta_{n  n'}\tilde Q_{ m'p }(t-\tau)\tilde Q_{p  m}(t)e^{{\rm i }\omega_{pn'}\tau}
C^*(\tau)
\Big]\\
&\qquad \quad-\Big[\tilde Q_{n  n'}(t-\tau) \tilde Q_{ m'm }(t)e^{-{\rm i }\omega_{nm'}\tau}C(\tau)+ \tilde Q_{n  n'}(t)
\tilde Q_{ m'm }(t-\tau)e^{{\rm i }\omega_{mn'}\tau}C^*(\tau)\Big]\Bigg\}e^{\ii k' \omega_{\rm d}(t-\tau)}\;.
\ee
Substituting the Fourier expansion~\eqref{result1} into
Eq.~\eqref{FR_tensor}, we obtain
\be{FmodesBMME2}
R^{k'(2)}_{nmn'm'}(t)
=&-\frac{1}{\hbar^2}\sum_{l ,l'}e^{\ii(l-l'+k')\omega_{\rm d} t}\int_{0}^\infty d\tau\Bigg\{\sum_{p}\Big[\delta_{mm'}Q_{np}^{-l'}Q^{l}_{p n'}e^{-{\rm i}
\omega_{p m'}^{l+k'} \tau}C(\tau)+ 
\delta_{nn'}Q_{m'p}^{-l'}Q^{l}_{p m}[e^{-{\rm i}
\omega_{p n'}^{l'-k'} \tau}C(\tau)]^*
\Big]\\
&
\qquad\qquad\qquad\qquad\qquad\qquad\qquad
- Q_{m'm}^{-l'}Q_{nn'}^{l}\left[e^{-{\rm i}
\omega_{nm'}^{l+k'} \tau}C(\tau)
+ [e^{-{\rm i}
\omega_{mn'}^{l'-k'} \tau}C(\tau)]^*\right]\Bigg\}\;.
\ee
Here, we have defined 
\[
\omega_{nm}^l:=\varepsilon_n - \varepsilon_m + l \omega_{\text d}\;.
\]
Consider now the Born-approximated version of the exact steady-state equation~\eqref{NZ_Fourier}, in which the kernel is replaced by Eq.~\eqref{FmodesBMME2}, namely by its leading (second) order truncation. Expanding in Fourier series the system RDM in the left-hand side of  Eq.~\eqref{NZ_Fourier} and matching the Fourier indexes, i.e. constraining  $l-l'+k'=k$ in Eq.~\eqref{FmodesBMME2}, yields the GFR equation for the coefficients of the expansion of $\tilde\rho(t)$ at the steady state
\be{FBME_Fourier}
0^{(k)}
=&-\ii\omega_{nm}^k\tilde\rho^{(k)}_{nm}+\sum_{n'm',k'}R_{nmn'm'}^{k,k'(2)}\tilde\rho^{(k')}_{n'm'}\;,
\ee
where the second-order GFR tensor reads
\be{FRTensor}
R_{nmn'm'}^{k,k'(2)}=&-\frac{1}{\hbar^2}\sum_{l}\Bigg\{\sum_{p }\Big[\delta_{mm'}Q_{n p }^{-l+k-k'}Q^{l}_{p n'}W_{p m'}^{l+k'}+ 
\delta_{nn'} Q_{m'p }^{-l+k-k'}Q^{l}_{p m}[W_{p n'}^{l-k}]^*\Big]\\
&\qquad\qquad\qquad -
 Q_{m'm}^{-l+k-k'}Q_{nn'}^{l}\left[W_{nm'}^{l+k'}
+ [W_{mn'}^{l-k} ]^*\right]\Bigg\}\;.
\ee
with
\be{WalphabetaSS}
W^{l}_{nm}\equiv W(\omega^{l}_{nm})=&\int_{0}^\infty d\tau\; e^{-{\rm i}
\omega_{nm}^l \tau}C(\tau)\;.
\ee
One can verify that Eq.~\eqref{FBME_Fourier} preserves the Hermiticity of the RDM
\be{}
\rho_{nm}^{(k)}=[\rho_{mn}^{(-k)}]^*\;.
\ee
\indent Besides the weak system-bath coupling (Born) approximation, no further approximations, such as Markovian or secular, were invoked in deriving the steady-state master equation~\eqref{FBME_Fourier}. Here, we see that the GFR tensor couples the different Fourier components of $\tilde\rho(t)$ as well as populations and coherences. This equation is used in the application to a driven dissipative qubit discussed in the following sections, as a benchmark for different approximated treatments. 

In the static limit $0=-\ii(\omega_n-\omega_m)\rho_{nm}^\infty+\sum_{n'm'}R_{nmn'm'}^{(2)}\rho_{n'm'}^\infty$, with 
\be{}
R_{nmn'm'}^{(2)}=&-\frac{1}{\hbar^2}\Bigg\{\sum_{p}\Big[\delta_{mm'}Q_{n p }Q_{p n'}W_{p m'}+ 
\delta_{nn'} Q_{m'p}Q_{p m} W_{p n'}^*\Big] -
 Q_{m'm}Q_{nn'}\Big[W_{nm'}
+W_{mn'}^*\Big]\Bigg\}\;,
\ee
namely, the Fourier (upper) indexes are all set to zero, the Floquet modes become the energy eigenstates, and the quasienergies become the eigenfrequencies. This is the steady-state version of what is referred to as the Redfield$^+$ equation in~\cite{Xu2021}.\\ 

\subsubsection{Steady-state heat current}

Starting from Eqs.~\eqref{Pb_exact}-\eqref{KI_second}, and using the expansion~\eqref{result1} for the system coupling operator, and assuming that the steady-state density matrix in the basis of Floquet modes is time-periodic,  we find for the steady state heat current, to leading order in the system-bath coupling,
\be{Pb_ss_Fourier}
P_{\rm b}(t)=&-\frac{1}{\hbar}\sum_{n ,n ',m'}\sum_{l,l',k'}\int_0^\infty d\tau\; 2{\rm Im}\Big\{e^{\ii(l-l'+k') \omega_{\text d} t}
Q_{m'n }^{-l'}Q_{n n '}^{l}
e^{-{\rm i}
\omega_{n m'}^{l+k'} \tau}\frac{d}{d\tau}C(\tau)
\tilde\rho_{n 'm'}^{(k')}\Big\}\\
=&-\frac{1}{\hbar}\sum_{n ,n ',m'}\sum_{l,l',k'} 2{\rm Re}\Big\{e^{\ii(l-l'+k') \omega_{\text d} t}
Q_{m'n }^{-l'}Q_{n n '}^{l}
\omega_{nm'}^{l+k'} W^{l+k'}_{nm'}
\tilde\rho_{n 'm'}^{(k')}\Big\}\;.
\ee
Here we used
\be{barW}
\int_{0}^\infty d\tau\; e^{-{\rm i}
\omega_{nm}^l \tau}\frac{d}{d\tau}C(\tau)
&=-C(0) +\ii \omega_{nm}^l W^{l}_{nm}\;.
\ee
Note that the term proportional to $C(0)$ in Eq.~\eqref{Pb_ss_Fourier} vanishes because of the sum over $n',m'$ and the symmetry of the integrand.
Fourier-expanding $P_{\rm b}(t)$ in Eq.~\eqref{Pb_ss_Fourier} gives for the Fourier components of the steady-state heat current to the bath
\be{Pb_k}
P_{\rm b}^{(k)}=&-\frac{1}{\hbar}\sum_{n ,n ',m'}\sum_{l,k'} \Big\{
Q_{m'n }^{-l+k-k'}Q_{n n '}^{l}
\omega_{nm'}^{l+k'} W^{l+k'}_{nm'}
\tilde\rho_{n 'm'}^{(k')}
+
[Q_{m'n }^{-l-k-k'}Q_{n n '}^{l}
\omega_{nm'}^{l+k'} W^{l+k'}_{nm'}
\tilde\rho_{n 'm'}^{(k')}]^*\Big\}%
\;,
\ee
where the Fourier components of $\tilde\rho(t\rightarrow \infty)$ are the solution of Eq.~\eqref{FBME_Fourier}.\\
\indent As in the GFR equation~\eqref{FBME_Fourier}, the only approximation involved in the formula for the steady-state heat current, Eq.~\eqref{Pb_k}, is weak system-bath coupling. No further Markovian or secular approximations are invoked. The one-period average of the power to the bath is given by the $k=0$ coefficient which explicitly reads
\be{Pb_k0}
P_{\rm b}^{(0)}=&-\frac{1}{\hbar}\sum_{n ,n ',m'}\sum_{l,k'} 2 {\rm Re}\Big\{
Q_{m'n }^{-l-k'}Q_{n n '}^{l}
\omega_{nm'}^{l+k'} W^{l+k'}_{nm'}
\tilde\rho_{n 'm'}^{(k')}\Big\}\;.
\ee

\subsection{Steady state in the Markovian approximation}
\label{ss_Markov}

As shown above, a time-local master equation for the Fourier components of the RDM and the corresponding heat current formula, Eqs.~\eqref{FBME_Fourier} and~\eqref{Pb_k}, can be obtained, at the steady state, without resorting to the Markovian approximation, namely by only invoking the weak coupling (Born) approximation.\\
\indent Here we give, for completeness, the corresponding formulas obtained by taking the steady-state limit of the Markovian master equation (Born-Markov), as detailed in Appendix~\ref{Markovian}. 
The Markovian, nonsecular Floquet-Redfield equation is formally similar  to the non-Markovian GFR equation~\eqref{FBME_Fourier}, namely
\be{FBME_FourierMarkovian}
0^{(k)}
=-\ii\omega_{nm}^k\tilde\rho^{(k)}_{nm}+\sum_{n'm',k'}R_{nmn'm'}^{k,k'(2)M}\tilde\rho^{(k')}_{n'm'}\;.
\ee
Here, $M$ stands for \emph{Markovian} and the Markovian Floquet-Redfield tensor reads, cf. Eq.~\eqref{FRTensor},
\be{FRTensorMarkovian}
R_{nmn'm'}^{k,k'(2)M}=&-\frac{1}{\hbar^2}\sum_{l}\Bigg\{\sum_{p }\Big[\delta_{mm'}Q_{n p }^{-l+k-k'}Q^{l}_{p n'}W_{p n'}^{l}+ 
\delta_{nn'} Q_{m'p }^{-l+k-k'}Q^{l}_{p m}[W_{p m'}^{l-k+k'}]^*\Big]\\
&\qquad\qquad\qquad -
 Q_{m'm}^{-l+k-k'}Q_{nn'}^{l}\left[W_{nn'}^{l}
+ [W_{mm'}^{l-k+k'} ]^*\right]\Bigg\}\;.
\ee
As Eq.~\eqref{FBME_Fourier}, also Eq.~\eqref{FBME_FourierMarkovian} preserves the Hermiticity of the RDM, $\rho_{nm}^{(k)}=[\rho_{mn}^{(-k)}]^*$.
The corresponding expression for the Fourier components of the heat current to bath $\rm b$ is, in the Markovian approximation,
\be{Pb_kMarkovian}
P_{\rm b}^{(k)M}=&-\frac{1}{\hbar}\sum_{n ,n ',m'}\sum_{l,k'} \Big\{
Q_{m'n }^{-l+k-k'}Q_{n n '}^{l}
\omega_{nn'}^{l} W^{l}_{nn'}
\tilde\rho_{n 'm'}^{(k')}
+
[Q_{m'n }^{-l-k-k'}Q_{n n '}^{l}
\omega_{nn'}^{l} W^{l}_{nn'}
\tilde\rho_{n 'm'}^{(k')}]^*\Big\}\;,
\ee
cf. Eq.~\eqref{Pb_k}, where the RDM elements $\tilde\rho_{n 'm'}^{(k')}$ are given by Eq.~\eqref{FBME_FourierMarkovian}.
The one-period average of the heat current to the bath, given by the $k=0$ coefficient, reads
\be{Pb_kMarkovian0}
P_{\rm b}^{(0)M}=&-\frac{1}{\hbar}\sum_{n ,n ',m'}\sum_{l,k'} 2{\rm Re} \Big\{
Q_{m'n }^{-l-k'}Q_{n n '}^{l}
\omega_{nn'}^{l} W^{l}_{nn'}
\tilde\rho_{n 'm'}^{(k')}
\Big\}\;,
\ee
In section~\ref{driven_dissipative_qubit} below, we compare this expression with the corresponding non-Markovian approach, Eq.~\eqref{Pb_k0}, in the case of a driven qubit. The results show that the heat current in the Markovian approximation deviates substantially from the non-Markovian approach for finite system-bath coupling $\alpha$.    

\subsection{Full secular master equation}
\label{FSMEss}

Let us apply to Eq.~\eqref{FBME_Fourier} the full secular approximation with respect to $\omega_{nm}^k$, assuming that these frequencies are much larger than the frequency scale $\gamma$ associated to the dissipator. This is the case if system-bath coupling is weak enough, in the absence of degeneracies. The steady-state populations in the Floquet eigenbasis (which coincide with the populations in the basis of Floquet modes), are then the solution of the master equation
\be{}
0=&-\frac{1}{\hbar^2}\sum_{n'}\sum_{l}\Bigg\{\sum_{p }\Big[\delta_{nn'}Q_{n p }^{-l}Q^{l}_{p n'}W_{p n'}^{l}+ 
\delta_{nn'} Q_{n'p }^{-l}Q^{l}_{p n}[W_{p n'}^{l}]^*\Big]\\
&\qquad\qquad\qquad\qquad\qquad -
\left[ Q_{n'n}^{-l}Q_{nn'}^{l}W_{nn'}^{l}
+ Q_{nn'}^{-l}
Q_{n'n}^{l}[W_{nn'}^{l} ]^*\right]\Bigg\}\rho_{n'n'}\;.
\ee
The equation above can be written as the following Pauli-type master equation for the populations in the Floquet basis 
\be{ME_ss_secular}
0=\sum_ m (\Gamma_{ n m}\rho_{m m}-\Gamma_{ m n}\rho_{n n})\;,
\ee
with the rates given by the sum over $l$ of the so-called partial rates 
\be{Gammal}
\Gamma_{ n m}&=\sum_l \Gamma_{ n m}^{(l)}=\frac{1}{\hbar^2}\sum_l |Q_{ n m}^l|^2 2{\rm Re} W_{ n m}^l \;.
\ee
Analogously, the one-period average of the steady-state heat current in the full secular approximation reads
\be{Pb_ss_secular}
P_{\rm b}^{(0)}=&-\frac{1}{\hbar}\sum_{n ,n '}\sum_{l} 2{\rm Re}\Big\{
Q_{n'n }^{-l}Q_{n n '}^{l}
\omega_{nn'}^{l} W^{l}_{nn'}
\rho_{n'n'}\Big\}\\
=&-\sum_{ n, m,l}\hbar\omega_{n m}^l\Gamma_{ n m}^{(l)}\rho_{m m}
\;,
\ee
where the steady-state populations $\rho_{nn}$ are the solution of Eq.~\eqref{ME_ss_secular}. These expressions largely simplify the calculation of the heat current with respect to the GFR approach at the cost of introducing the assumption $\omega_{nm}^k\gg \gamma$, where $\gamma$ is the frequency scale of the dissipator, which can become questionable for certain drive parameters~\cite{Hotz2021}.
The impact of neglecting nonsecular terms is discussed using a numerically exact method in~\cite{Shen2014} and with a generalized, Markovian master equation in~\cite{Akbari2026}. Moreover, the full secular master equation is shown to be thermodynamically inconsistent in~\cite{Tude2026}, as it predicts zero work performed by the drive (as a result of vanishing steady-state coherences in the Floquet eigenbasis) given a finite steady-state heat current.\\
\indent Equation~\eqref{ME_ss_secular} and the corresponding heat current formula, Eq.~\eqref{Pb_ss_secular}, also derived using different methods in~\cite{Langemeyer2014, Gasparinetti2014}, were used to model the heat current to an engineered bath in a driven superconducting qubit-resonator system experiment in~\cite{Satrya2026}.  In the next section, we apply the secular master equation approach to the heat transport in the driven spin-boson model and compare its results with those of the GFR equation~\eqref{FBME_Fourier}, which is used as a benchmark.  

\section{Application: Steady-state heat transport in a driven qubit}
\label{driven_dissipative_qubit}

\begin{figure}[ht!]
\begin{center}
\includegraphics[width=0.89\textwidth,angle=0]{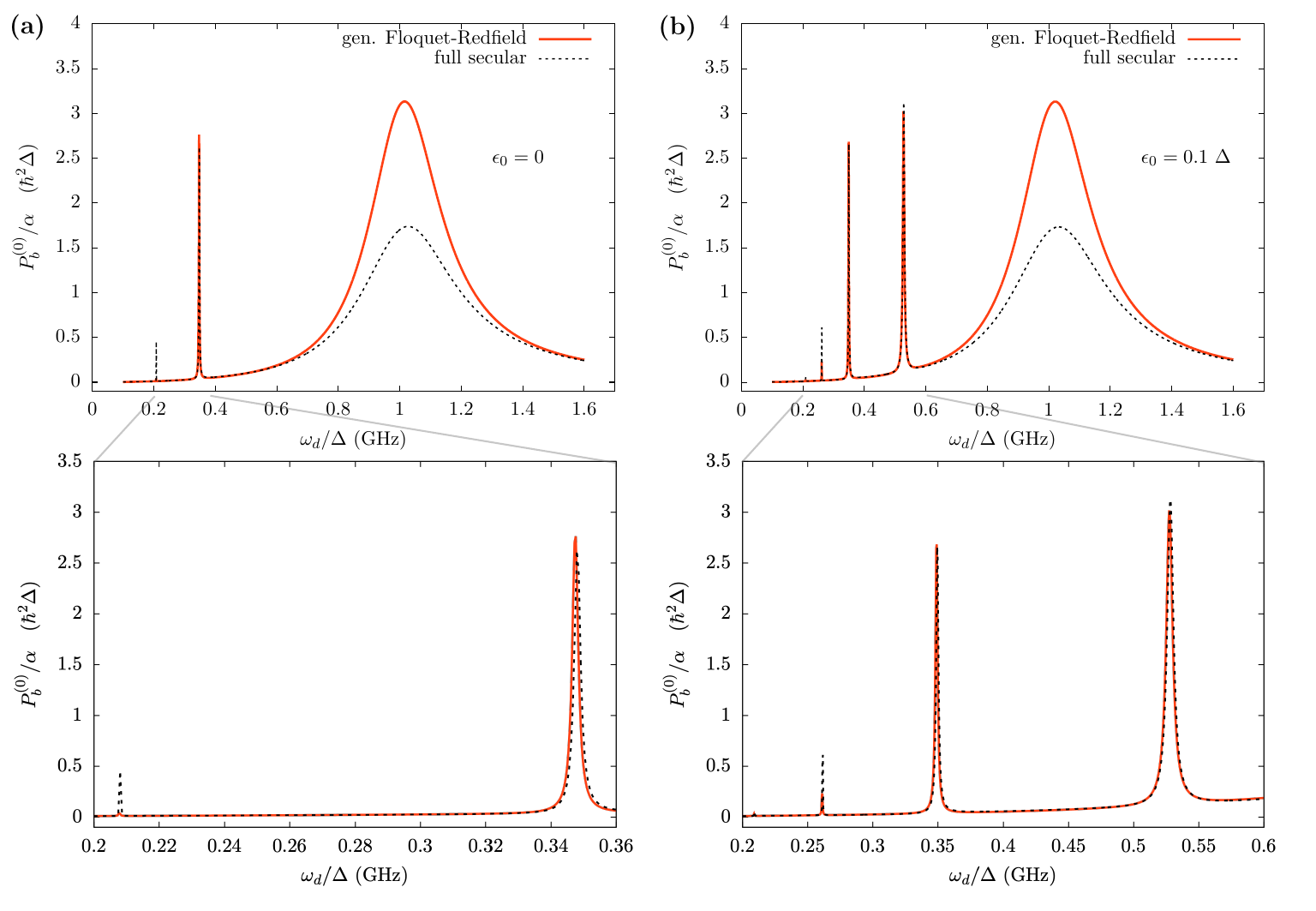}
\caption{Driven qubit. Steady-state heat current to the bath \emph{vs.} drive frequency for $A_{\rm d}=0.4~\Delta$, at zero and finite static bias. Details of the multi-photon peaks in the lower panels. Red curves: Numerical solution of the steady-state generalized Floquet-Redfield master equation~\eqref{FBME_Fourier} and Eq.~\eqref{Pb_k0} (one-period average of the heat current to the bath). The only approximation used is weak system-bath coupling. Dashed, black lines: Numerical solution of the full secular master equation and heat current,  
Eqs.~\eqref{ME_ss_secular}-\eqref{Pb_ss_secular}. The bath has an ohmic-Drude spectral density function, Eq.~\eqref{G_Drude}, with cutoff $\omega_c=20~\Delta$ and coupling strength $\alpha=0.0003$ (relevant for the full  Floquet-Redfield master equation), and temperature $T=0.3~\hbar\Delta/k_B$. The truncation of the sums over $l\in [-L,L]$ and $k'\in [-K,K]$ in the Redfield tensor and in Eq.~\eqref{Pb_k} are $K=6$ and $L=20$, while is the number of terms retained in the sum over the Matsubara frequencies, Eq.~\eqref{W_omega_Drude}, is $N_{\rm Mats}=200$.} 
\label{simulationTLSA04}
\end{center}
\end{figure}

In Fig.~\ref{simulationTLSA04}, we show the results for the zero-frequency steady-state heat current $P_{\rm b}^{(0)}$ to the bath from the GFR equation (Eqs.~\eqref{FBME_Fourier} and~\eqref{Pb_k}) and compare them with the ones from the full secular master equation approach, Eqs.~\eqref{ME_ss_secular} and~\eqref{Pb_ss_secular}.\\
\indent To be definite, we specialize the equations to a driven qubit with Hamiltonian
\be{}
\hat{H}_{\rm qb}(t)=-\frac{\hbar}{2}\left[\Delta\hat \sigma_x + \epsilon(t)\hat \sigma_z \right]\;,\qquad \epsilon(t)=\epsilon_0 + A_{\text d}\sin(\omega_{\text d} t)\;,
\ee
given in the so-called qubit localized basis $\{\ket{\pm}\}$ of eigenstates of $\hat\sigma_z=\ket{+}\bra{+}-\ket{-}\bra{-}$\;. The qubit interacts with a single heat bath according to Eq.~\eqref{Hterms} with $\hat Q=\hat\sigma_z$. As a result, the counterterm renormalizing the system Hamiltonian according to the Caldeira-Leggett prescription, see Eq~\eqref{A2} in Appendix~\ref{CaldeiraLeggett_model}, produces a constant energy shift and can be ignored altogether in the present application. The bath is characterized by the ohmic-Drude spectral density function 
\be{G_Drude}
G(\omega)=\frac{\alpha\omega}{1+(\omega/\omega_c)^2}\;,
\ee
see also Eq.~\eqref{W_omega_Drude}. This driven qubit-plus-bath model, is referred to as driven spin-boson model in the literature~\cite{Weiss2012}.\\
With our conventions, the coupling parameter $\alpha$ is related to the conventional ohmic spin-boson dissipation strength, or Kondo parameter $\alpha_{\rm K}$, by $\alpha=\alpha_{\rm K}/2$. The zero-temperature localization transition~\cite{Weiss2012} in the undriven case (scaling limit $\omega_c\gg \Delta$) occurs therefore at $\alpha=1/2$ with our conventions.\\
\indent The plots in Fig.~\ref{simulationTLSA04} display peaks of the heat current to the bath at drive frequencies that roughly match fractions of the one-period average $\bar\omega_{\rm q}$ of the qubit frequency $\omega_{\rm q}(t)$, namely $\omega_{\rm d}\simeq \bar\omega_{\rm q}/l$. Moreover, the even fractional peaks, $l=2,4,..$, are suppressed in the symmetric case, i.e. at zero static bias $\epsilon_0$, which reflects the presence of a selection rule holding at the qubit symmetry point~\cite{Grifoni1995, Ho1986, Shytov2003, Shevchenko2005, Deppe2008, Thomas2023}. An intuitive account of this is provided in Sec.~\ref{selection_rule} below. The width and height of the peaks decreases with $l$ and we notice that the height of the low-frequency fractional peaks tend to be overestimated by the full secular master equation as compared to the GFR equation. Besides these low-frequency behavior, a resonant, broad peak is found at $\omega_{\rm d}\simeq \bar\omega_{\rm q}$. In this case, the full secular approach largely underestimates the peak-value, giving roughly half of the value predicted by the nonsecular GFR approach.     
\begin{figure}[ht!]
\begin{center}
\includegraphics[width=0.45\textwidth,angle=0]{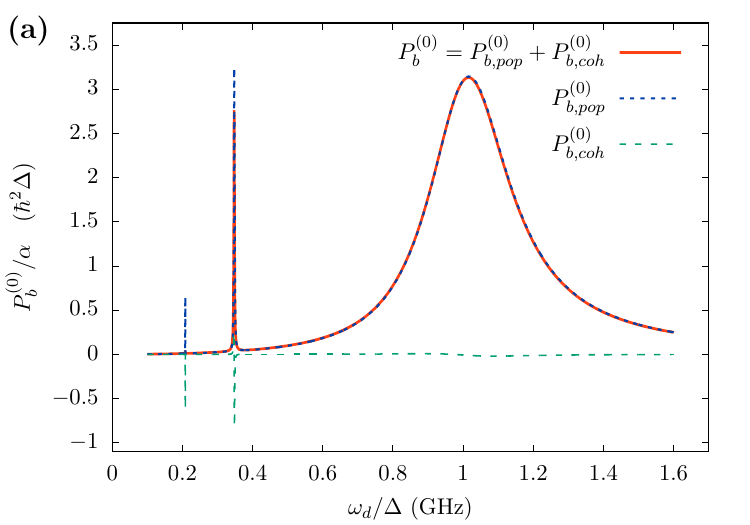}
\includegraphics[width=0.45\textwidth,angle=0]{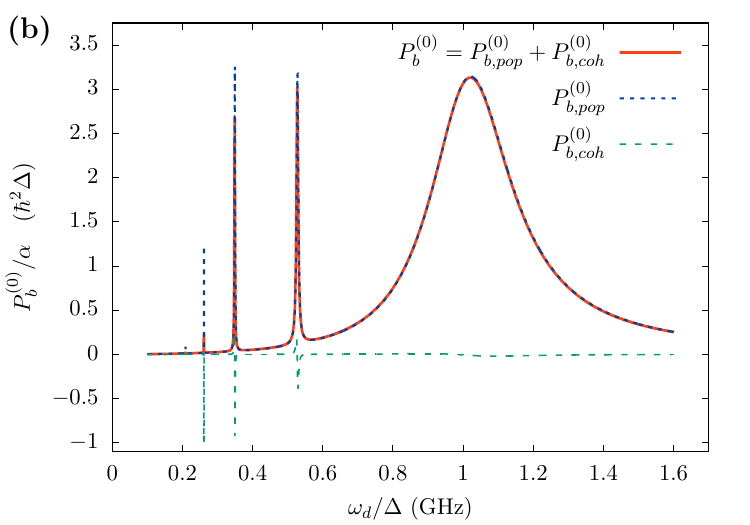}
\caption{Impact of steady-state coherences in the generalized Floquet-Redfield (GFR) approach. Steady-state heat current to the bath \emph{vs.} drive frequency for $A_{\rm d}=0.4~\Delta$, at zero (left) and finite (right) static bias, $\epsilon_0=0,0.1~\Delta$. Orange, solid lines: Numerical solution of the GFR equation~\eqref{FBME_Fourier} and Eq.~\eqref{Pb_k0} (one-period average of the heat current). Dashed, blue and red lines: separate contributions to $P_{\rm b}^{(0)}$ from populations and coherences, respectively, according to Eq.~\eqref{Pb_0_contributions}. Parameters are as in Fig.~\ref{simulationTLSA04}} 
\label{simulationTLSA04_contributions}
\end{center}
\end{figure}

To gain insight in the different low-frequency behavior between the two approaches for calculating the heat current $P_{\rm b}$ to the bath, in Fig~\ref{simulationTLSA04_contributions} we show the result from the GFR equation, with the contribution to the steady-state heat current $P_{\rm b}^{(0)}=P_{\rm b,\;pop}^{(0)}+P_{\rm b,\;coh}^{(0)}$ from the  populations separated from that of the  coherences, namely
\be{Pb_0_contributions}
P_{\rm b\;pop}^{(0)}=&-\frac{1}{\hbar}\sum_{n ,m}\sum_{l, k'} 2{\rm Re}\Big\{
Q_{mn }^{-l-k'}Q_{n m}^{l}
\omega_{n m}^{l+k'} W^{l+k'}_{n m}
\tilde\rho_{mm}^{(k')}\Big\}\;,\\
P_{\rm b\;coh}^{(0)}=&-\frac{1}{\hbar}\sum_{n ,n ',m'(\neq n')}\sum_{l,k'} 2{\rm Re}\Big\{
Q_{m'n }^{-l+k-k'}Q_{n n '}^{l}
\omega_{nm'}^{l+k'} W^{l+k'}_{nm'}
\tilde\rho_{n 'm'}^{(k')}\Big\}\;.
\ee
Note that, fixing $k'=0$ in the first of Eqs.~\eqref{Pb_0_contributions} reproduces the full secular version, Eq.~\eqref{Pb_ss_secular}, but does not give the same result as the full secular approach. This is because the zero-frequency component of the populations here are still the solutions of the GFR equation which couples them to the steady-state coherences and to the different Fourier components $k$.\\
\indent From Fig.~\ref{simulationTLSA04_contributions} emerges that the steady-state coherences impact the low-frequency peaks reducing their height, with respect to the one obtained from the populations alone. On the other hand, the resonant maximum is barely influenced by the steady-state coherences and essentially rendered by $P_{\rm b\;pop}^{(0)}$. This means that, in this resonant frequency regime, higher harmonics of the populations, obtained from the Floquet-Redfield equation, play an important role, as well as the different sidebands with $k'\neq 0$ involved in $\omega^{l+k'}_{nm}$ in the  formula for $P_{\rm b\;pop}^{(0)}$.

\begin{figure}[ht!]
\begin{center}
\includegraphics[width=0.297\textwidth,angle=0]{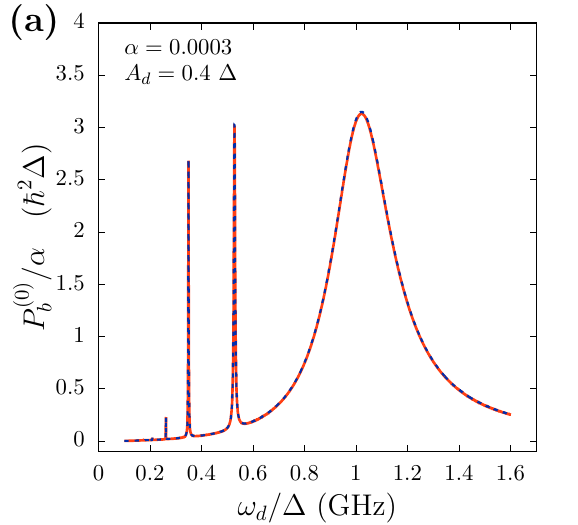}
\hspace{-0.85cm}
\includegraphics[width=0.27\textwidth,angle=0]{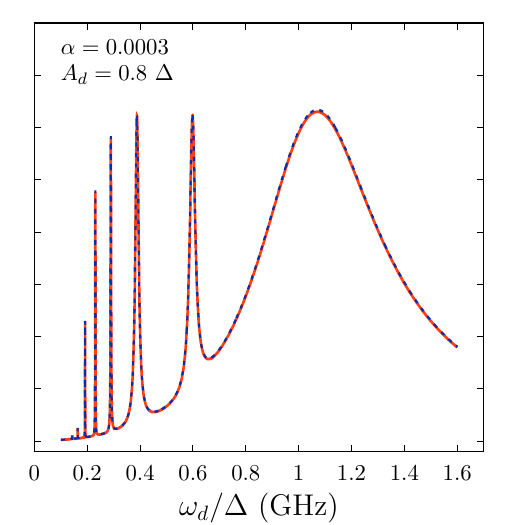}
\hspace{-0.85cm}
\includegraphics[width=0.27\textwidth,angle=0]{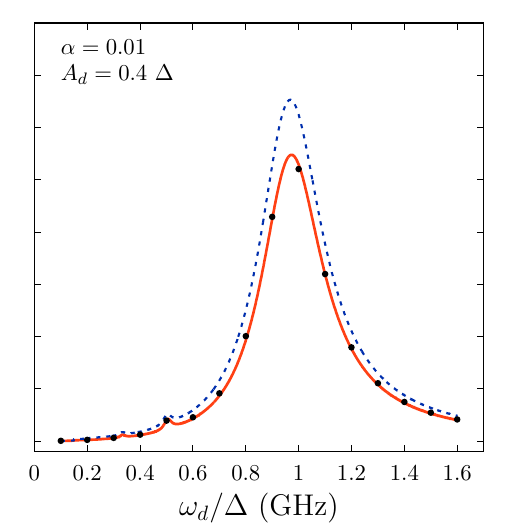}
\hspace{-0.85cm}
\includegraphics[width=0.27\textwidth,angle=0]{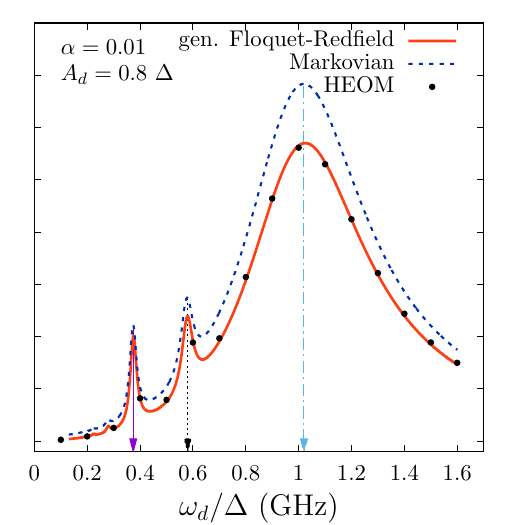}
\includegraphics[width=0.543\textwidth,angle=0]{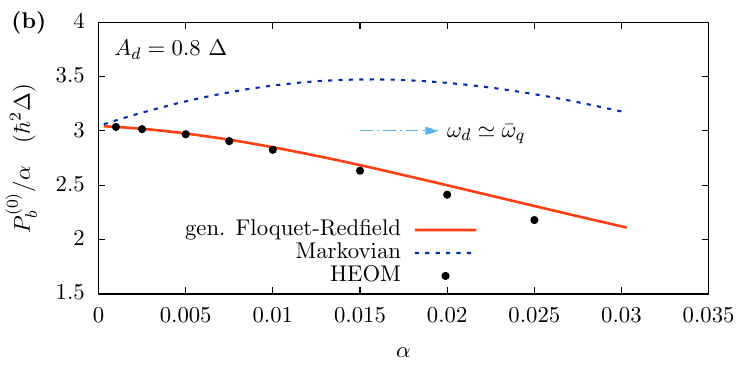}
\hspace{-0.8cm}
\includegraphics[width=0.49\textwidth,angle=0]{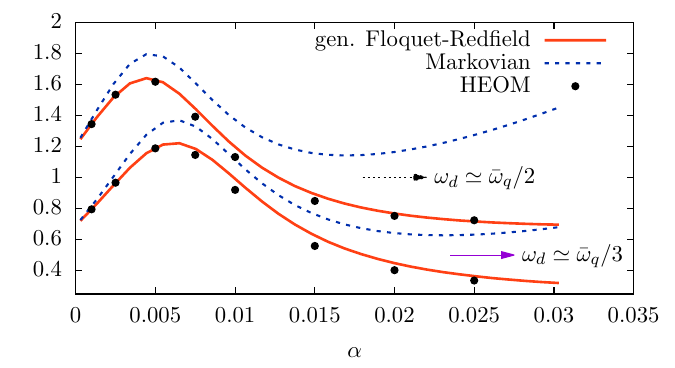}
\caption{Comparison between the non-Markovian generalized Floquet-Redfield (GFR) equation and the Markovian Floquet-Redfield equation approach at the steady state for finite static bias, $\epsilon_0=0.1~\Delta$. Red, solid lines: GFR  equation, Eqs.~\eqref{FBME_Fourier} and~\eqref{Pb_k0} (one-period average of the heat current). Dashed, dark-blue lines: Markovian Floquet-Redfield master equation  
Eqs.~\eqref{FBME_FourierMarkovian}-\eqref{Pb_kMarkovian0}. At very weak system-bath coupling $\alpha$, the two approaches give overlapping results while at larger $\alpha$ the heat current differs noticeably. (a) -- Zero-frequency, steady-state heat current to the bath \emph{vs.} drive frequency for four combinations of system-bath coupling $\alpha$ and drive amplitude $A_{\rm d}$. (b) -- Zero-frequency, steady-state heat current \emph{vs.}  system-bath coupling $\alpha$ at $A_{\rm d}=0.8~\Delta$ for three drive frequencies corresponding to the three heat current peaks highlighted in panel (a) namely $\omega_{\rm d}/\Delta = 0.3735,\,0.58,\,1.0197$. Note that the heat current is scaled with $\alpha$. Other parameters are as in Fig.~\ref{simulationTLSA04} with the following exception: for $A_{\rm d}=0.8~\Delta$, we set $K=10$ and $L=30$.} 
\label{simulationTLSMarkovian}
\end{center}
\end{figure}

\indent Next, we asses the the impact of the Markovian approximation, discussed in the general case in Sec.~\ref{ss_Markov} and Appendix~\ref{Markovian}, on the heat current calculation for the driven spin-boson model at the steady state considered here. This approximation is usually employed in master equation approaches as a starting point for evaluations or further approximations, eg. the secular approximation which gives a master equation for the RDM in Lindblad form.\\
\indent In Fig.~\ref{simulationTLSMarkovian}, the zero-frequency heat current to the bath $P_{\rm b}^{(0)M}$ in the Markovian approximation, obtained from Eqs.~\eqref{FBME_FourierMarkovian}-\eqref{Pb_kMarkovian}, is compared to the same quantity calculated using the non-Markovian GFR approach, Eqs.~\eqref{FBME_Fourier}-\eqref{Pb_k}, at different coupling strengths $\alpha$ and drive amplitudes. The results show excellent agreement for the value of the system-bath coupling $\alpha=0.0003$ chosen in Fig~\ref{simulationTLSA04} and in the evaluations made in the following sections. However, increasing $\alpha$, the results of the time-local Floquet-Redfield approach deviate noticeably from those of the GFR approach, even for the chosen ohmic spectral density function, Eq.~\eqref{G_Drude}.

To assess if the GFR equation constitutes a genuine improvement over the Markovian, nonsecular Floquet-Redfield approach, we include simulations using the nonperturbative HEOM technique~\cite{Tanimura1989,Tanimura2020,Kato2016}. The results show that the GFR performs remarkably well beyond the vanishing coupling limit, where its prediction depart from those of the Markovian theory. 
A similar agreement between the non-Markovian Born equation and the HEOM is highlighted in~\cite{Xu2021} in the static case.
We also note that the heat current as a function of $\alpha$, shown at resonance in panels (b) of Fig.~\ref{simulationTLSMarkovian}, displays a nonlinear behavior in both Markovian and the non-Markovian case. This is a feature of the nonsecular approach, where $P^{(0)}_{\rm b}(\alpha)$ is not just proportional to $\alpha$, as in the full secular approach, see Sec.~\ref{FSMEss}. Indeed, the nonsecular equations for the RDM,  feature \emph{unitary} evolution terms, containing the extended quasi-energies $\omega_{nm}^k$ independent of $\alpha$, coupled to the dissipative part, which depend on $\alpha$.\\ 
The nonlinearity of the heat current as a function of $\alpha$ reported here is therefore an exquisitely leading-order (Born approximation) feature and does not account for higher-order corrections in the system-bath coupling. Higher-order master equation approaches show that the leading-order, Markovian approach fails in the static case with sub-ohmic bath~\cite{Crowder2024} and is insufficient, without higher-order contributions, to reproduce the correct equilibrium mean-force Gibbs state~\cite{Kumar2025}.\\
The nonlinear behavior as a function of $\alpha$ shown in Fig.~\ref{simulationTLSMarkovian}(b) is explicitly derived in Sec.~\ref{comparison_analytical_TLS} below (see Eq.~\eqref{Pb_vs_alpha_analytical}) where we find an analytical approximate solution for the steady-state heat current in the driven spin-boson model within a leading order, Markovian approach.

\section{Transient within the Markovian approximation}

To address the transient dynamics, and in view of the approach used in Sec.~\ref{inst_eigenbasis}, we start from the Nakajima-Zwanzig equation~\eqref{NZ_equation} with the perturbative, weak-coupling expression~\eqref{NZ_kernel_second} for the kernel, and perform the Markov approximation in the interaction picture. The resulting master equation is then expressed in the basis of Floquet modes, obtaining a Bloch-Redfield-type master equation. On top of the Markovian approximation, we also assume that the bath correlation function $C(t)$, Eq.~\eqref{C_t}, decays fast as compared to the time scale of the drive.
Following the procedure detailed in Appendix~\ref{Markovian}, see Eq.~\eqref{A_FRME} therein, we end up with the following time-local master equation with time-dependent Floquet-Redfield tensor which is best suited for the regime of slow driving
\be{FBMME}
\dot{\tilde\rho}_{nm}(t) \simeq
 -\ii\omega_{nm}\tilde\rho_{nm}(t) -\frac{1}{\hbar^2}
\sum_{n'm'}&\Big\{\sum_p\Big[\delta_{mm'}\tilde Q_{np}(t) \tilde Q_{pn'}(t)W_{pn'} + \delta_{nn'}\tilde Q_{m'p}(t)\tilde Q_{pm}(t)W_{pm'}\Big]\\ 
&\qquad \qquad -
\tilde Q_{m'm}(t)\tilde Q_{nn'}(t)\Big[W_{nn'} + W_{mm'}^*\Big]\Big\}\tilde\rho_{n'm'}(t)\\
\equiv -\ii\omega_{nm}\tilde\rho_{nm}(t)+\sum_{n'm'}&R^{(2)M}_{nmn'm'}(t)\tilde\rho_{n'm'}(t)
\;.
\ee

\indent Under the same approximations, we find for the heat current to the bath, see Appendix~\ref{Markovian}, 
\be{PbMarkov}
P_{\rm b}(t)\simeq 
&-\sum_{n ,n',m'}\hbar \omega_{nn'}2{\rm Re}\Big\{ 
\tilde Q_{m'n }(t)\tilde Q_{n n '}(t)\tilde\rho_{n 'm'}(t)W_{nn'}/\hbar^2\Big\}
\;,
\ee
where we applied the Markovian approximation and used Eq.~\eqref{barW}. Here, the elements $\tilde\rho_{n 'm'}(t)$ of the RDM in the basis of Floquet modes  are the solution of Eq.~\eqref{FBMME}.

\subsection{Transient dynamics of a driven two-level system within the secular approximation}

For well-separated quasienergies, we neglect, in the Markovian master equation in the basis of Floquet states, see Eq.~\eqref{A_FBMME}, the terms with fast oscillating complex exponentials. The elements of the Floquet-Redfield tensor in Eq.~\eqref{FBMME} which are retained are thus only $R^{(2)}_{nnmm}(t)$ and $R^{(2)}_{nmnm}(t)$. Consistently, in the heat current formula, this amounts to setting $n'=m'$, meaning that only the populations contribute to the heat current. For a two-level system, this translates into the following secular master equation in the Floquet basis
\be{METLS}
\partial_t\rho_{00}(t)&=\Gamma_\downarrow(t)-\Gamma_\Sigma(t)\rho_{00}(t)\\
\partial_t\rho_{01}(t) &=R^{{\rm M}(2)}_{0101}(t)\rho_{01}(t)\;.
\ee
Here, a remark on the secular approximation is in order. This approximation is applied to Eq.~\eqref{FBMME} with respect to the quasienergy difference $\omega_{nm}$, rather than separately on $\omega_{nm}$ and on the sideband index $k$, as done in the full secular, steady-state master equation~\eqref{ME_ss_secular}. The latter version, also called ultra-secular approximation in~\cite{Hotz2021}, is more restrictive, as one could have quasi-degenerate frequencies $\omega_{nm}^k$ even for well-separated quasienergies. Note that in the static case the two approaches coincide.\\
\indent For the heat current to the bath, we have similarly 
\be{PbTLSsecular}
P_{\rm b}(t)&=\hbar\omega_{10}\Gamma_\downarrow(t)\rho_{11}(t)-\hbar\omega_{10}\Gamma_\uparrow(t)\rho_{00}(t)\\
&=\hbar\omega_{10}\Big[\Gamma_\downarrow(t)-\Gamma_\Sigma(t)\rho_{00}(t)\Big]\;.
\ee
Here, $\Gamma_\Sigma(t)=\Gamma_\uparrow(t)+\Gamma_\downarrow(t)$. The rates are given by ($W_{00}=W_{11}=W$)
\be{ratesTLS}
\Gamma_\downarrow(t)&=R^{{\rm M}(2)}_{0011}(t)=|\tilde Q_{01}(t)|^2 2{\rm Re} W_{01}/\hbar^2\\
\Gamma_\uparrow(t)&=R^{{\rm M}(2)}_{1100}(t)=-R^{{\rm M}(2)}_{0000}(t)=|\tilde Q_{10}(t)|^2 2{\rm Re} W_{10}/\hbar^2\\
R^{{\rm M}(2)}_{0101}(t)&=-[\tilde Q_{00}(t) -\tilde Q_{11}(t)]^2W/\hbar^2 - |\tilde Q_{01}(t)|^2 [W_{10}+W_{01}^*]/\hbar^2\;.
\ee
If $\tilde Q_{nn}(t)=0$, or we can neglect $W'=\lim_{\omega_{nm}\to 0}W'_{nm} \propto \alpha/\beta$, see Eq.~\eqref{Re_W_omega}, namely neglecting pure dephasing,  then $R^{{\rm M}(2)}_{0101}(t)=-\Gamma_\Sigma(t)/2+\ii \gamma(t)$.

\section{Heat transport in the driven spin-boson model: Master equation in the instantaneous eigenbasis}
\label{inst_eigenbasis}

In this section we consider an alternative approach, initiated by Davies and Spohn~\cite{Davies1978} and developed in~\cite{Pekola2016,Karimi2016,KarimiPhDthesis} in the context of quantum thermodynamics with superconducting circuits. It is based on a time-dependent master equation in the instantaneous eigenbasis of a driven two-level system coupled to a heat bath. Its performance is assessed by comparing it to the results of the GFR approach.\\
\indent For the driven, dissipative qubit described in Sec.~\ref{driven_dissipative_qubit}, the instantaneous qubit frequency is
\be{}
\omega_{\text q}(t)=\sqrt{\Delta^2 + \epsilon^2(t)}\;.
\ee
Consider the basis states
\be{Fstates}
\ket{\psi_{0/1}(t)} = e^{\pm \ii\varphi(t)/2}\ket{\phi_{0/1}(t)}\;,\quad {\rm where}\quad \varphi(t) = \int^{t}_0 dt' \omega_{\text q}(t')\;.
\ee
Here, the functions $\ket{\phi_{0/1}(t)}$ follow instantaneously the qubit energy levels, $\hat{H}_{\rm qb}(t)\ket{\phi_{0/1}(t)}=\mp (\hbar \omega_{\rm q}(t)/2)\ket{\phi_{0/1}(t)}$. They read 
\begin{equation}\label{Fmodes}
\begin{aligned}
\ket{\phi_0(t)}=&\;c_-(t)\ket{-} + 
c_+(t)\ket{+}\\
\ket{\phi_1(t)}=&\;c_+(t)\ket{-} - 
c_-(t)\ket{+}
\end{aligned}\,,
\qquad
c_\pm (t)= \sqrt{\frac{ \omega_{\text q}(t) \pm \epsilon(t)}{2 \omega_{\text q}(t)}}\;.
\end{equation}
Using 
\[
\dot c_\pm(t)=\pm \frac{\Delta^2\dot\epsilon(t)}{4 \omega_{\rm q}^3(t)}\frac{1}{c_\pm(t)}=\pm \frac{\Delta \dot\epsilon(t)}{2 \omega_{\rm q}^2(t)} \;c_\mp(t)\;,
\]
entails 
\be{tderivatives}
\partial_t \ket{\psi_0(t)}=&\;\ii\frac{\dot\varphi(t)}{2}\ket{\psi_0(t)}-\frac{\Delta \dot\epsilon(t)}{2 \omega_{\rm q}^2(t)} e^{\ii\varphi(t)} \ket{\psi_1(t)}
=\ii\frac{\omega_{\rm q}(t)}{2}\ket{\psi_0(t)}-\frac{a}{2}\frac{\Delta\cos(\omega_{\rm d}t)}{1+[\epsilon(t)/\Delta]^2}e^{\ii\varphi(t)}\ket{\psi_1(t)}\\ \partial_t\ket{\psi_1(t)}=&-\ii\frac{\dot\varphi(t)}{2}\ket{\psi_1(t)}+\frac{\Delta \dot\epsilon(t)}{2 \omega_{\rm q}^2(t)} e^{-\ii\varphi(t)} \ket{\psi_0(t)}
=-\ii\frac{\omega_{\rm q}(t)}{2}\ket{\psi_1(t)}+\frac{a}{2}\frac{\Delta\cos(\omega_{\rm d}t)}{1+[\epsilon(t)/\Delta]^2}e^{-\ii\varphi(t)} \ket{\psi_0(t)}\;,
\ee
where we introduced the parameter $a=\omega_{\rm d}A_{\rm d}/\Delta^2=(4\delta)^{-1}$, where $\delta$ is the so-called adiabaticity parameter~\cite{Ivakhnenko2023}. 
If we neglect the terms proportional to $a$,  (in the adiabatic limit $a\ll 1$) the states $\ket{\psi_{0/1}(t)}$ satisfy the Schrödinger equation at every $t$ and are therefore Floquet states of the driven system, $\ket{\psi_{0/1}(t)}\simeq \ket{\Psi_{0/1}(t)}$.

For the matrix elements of the RDM in the basis~\eqref{Fstates} we have
\[
\partial_t \rho_{nm}(t)=
\bra{\psi_n(t)}\dot \rho(t)\ket{\psi_m(t)}+\frac{\ii}{\hbar}\bra{\psi_n(t)}[\hat{H}_{\rm qb}(t),\rho(t)]\ket{\psi_m(t)}+  \dots\;,
\]
where the dots denote the additional terms in Eq.~\eqref{tderivatives}. Note that, if the states $\ket{\Psi_n (t)}$ solve the Schr\"odinger equation, then these additional contributions vanish. 
Using the same dissipator as in Eq.~\eqref{METLS}, we arrive at the following master equation in the instantaneous eigenbasis~\eqref{Fstates}~\cite{Pekola2016,Karimi2016,KarimiPhDthesis}
\be{MEad}
\partial_t \rho_{00}(t)=& -{\frac{\dot q(t)}{1+q^2(t)}} \mathfrak{Im}\left(e^{\ii\varphi(t)} \rho_{01}(t)\right)-\Gamma_\Sigma (t)\rho_{00}(t)+\Gamma_\downarrow(t)\;,\\
\partial_t \rho_{01}(t)=& \; \ii {\frac{\dot q(t)}{1+q^2(t)}} e^{-i\varphi(t)} \left(\rho_{00}(t)-\frac{1}{2}\right) -\frac{1}{2}\Gamma_\Sigma(t) \rho_{01}(t)\;,
\ee
where $q(t)=\epsilon(t)/\Delta$ and $\Gamma_\Sigma = \Gamma_\uparrow + \Gamma_\downarrow$.\\
\indent 
Equation.~\eqref{MEad} couples coherences and populations via the Berry connection terms, see~\cite{Ilinskaya2026} for an analysis of the different basis choices when applying a master equation in Lindblad form to a driven dissipative system. 
However, the rates of the master equation are nontrivial to calculate in the basis~\eqref{Fstates}. It is therefore convenient to have them in the same form,  Eq.~\eqref{METLS}, as in the Floquet basis. To this end, we assume that, in the dissipator, the propagation $U_{\rm S}(t_2,t_1)|\Psi_n(t_1)\rangle=|\Psi_n(t_2)\rangle$ holds for the instantaneous eigenstates~\eqref{Fstates} or, equivalently, we assume $U_{\rm S}(t,t-\tau)\simeq \exp[-\ii\hat H_{\rm S}(t)\tau/\hbar]$, and then we apply the secular approximation. This is the case, however, only in the adiabatic limit $a \ll 1$, where Eq.~\eqref{MEad} coincides with the secular master equation in the basis of Floquet states, Eq.~\eqref{METLS}. In the general case, using the same rates as in Eq.~\eqref{METLS} constitutes an inconsistency for finite $a$, because the unitary evolution in Eq.~\eqref{MEad} is exact. Nevertheless, for the purpose of demonstrating the qualitative behavior of the driven qubit, this choice has the merit of not requiring the calculation of the Floquet eigensystem and turns out to provide excellent results for small enough $a$. Indeed, it properly describes the multi-photon peaks at fractional frequencies $\omega_{\rm d}=\bar\omega_{\rm q}/l$ even for moderately strong drive, see Fig.~\ref{simulationTLS_instantaneous} below.\\
\indent The matrix elements of the system coupling operator entering the rates of the master equation in the basis~\eqref{Fstates} read
\be{matrix_element}
|Q_{01}(t)|^2 &=|\bra{\psi_0(t)}\hat\sigma_z\ket{\psi_1(t)}|^2=\Bigg|\frac{-\Delta e^{-\ii\varphi(t)}}{\sqrt{\Delta^2+\epsilon^2(t)}}\Bigg|^2=\frac{\Delta^2}{\Delta^2+\epsilon^2(t)}\\
Q_{00}(t) &=\bra{\psi_0(t)}\hat\sigma_z\ket{\psi_0(t)}=-\frac{\epsilon(t)}{\sqrt{\Delta^2+\epsilon^2(t)}} =-Q_{11}(t)
\;.
\ee
Comparing Eq.~\eqref{MEad} with Eq.~\eqref{METLS}, one can see that we have approximated $R_{0101}(t)\simeq -\Gamma_\Sigma(t)/2$, neglecting the imaginary part of $R_{0101}(t)$ and $W_{nn}'$.

Similarly the heat current to the bath, in the adiabatic limit, is
\begin{equation}
\label{power}
P_{\rm b}(t)=\hbar \omega_{\rm q}(t)[\Gamma_\downarrow(t) -\Gamma_\Sigma(t) \rho_{00}(t)]\;,
\end{equation}
cf. Eq.~\eqref{PbTLSsecular}.

\begin{figure}[ht!]
\begin{center}
\includegraphics[width=0.48\textwidth,angle=0]{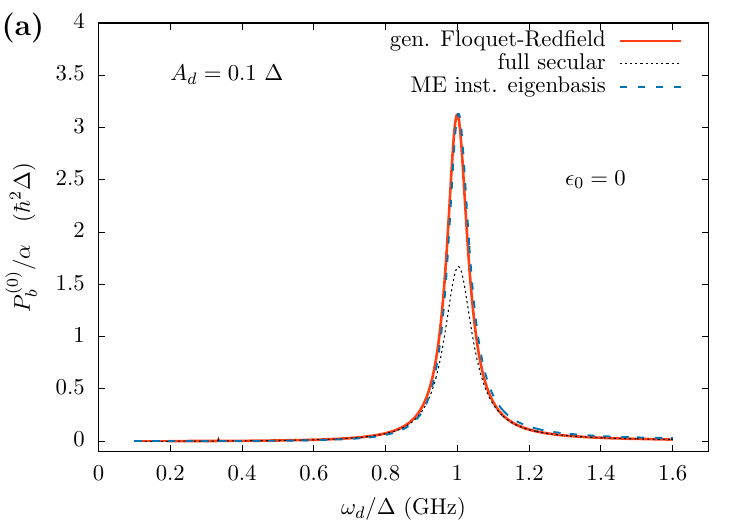}
\includegraphics[width=0.48\textwidth,angle=0]{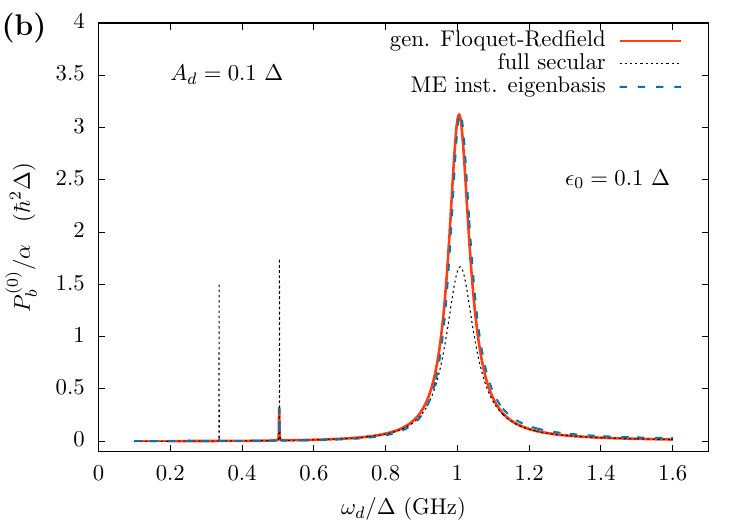}
\vspace{0.1cm}
\includegraphics[width=0.48\textwidth,angle=0]{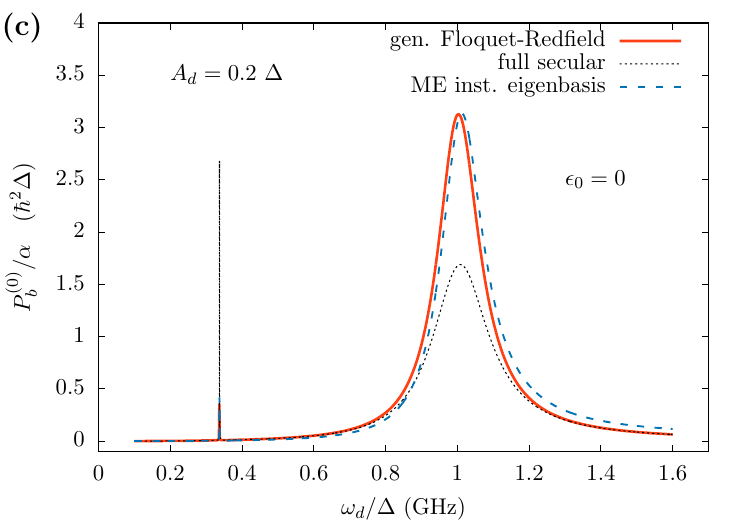}
\includegraphics[width=0.48\textwidth,angle=0]{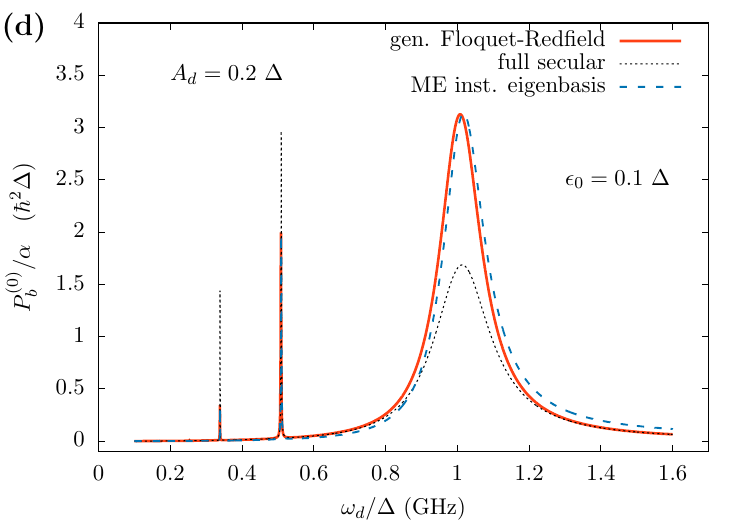}
\vspace{0.1cm}
\includegraphics[width=0.48\textwidth,angle=0]{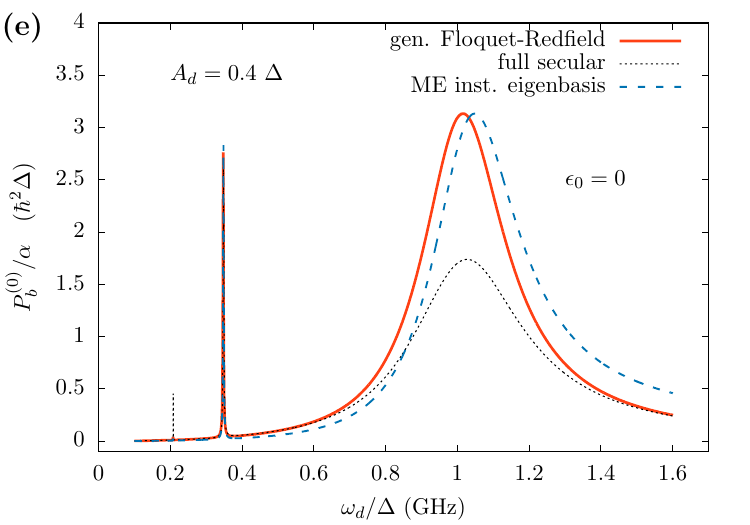}
\includegraphics[width=0.48\textwidth,angle=0]{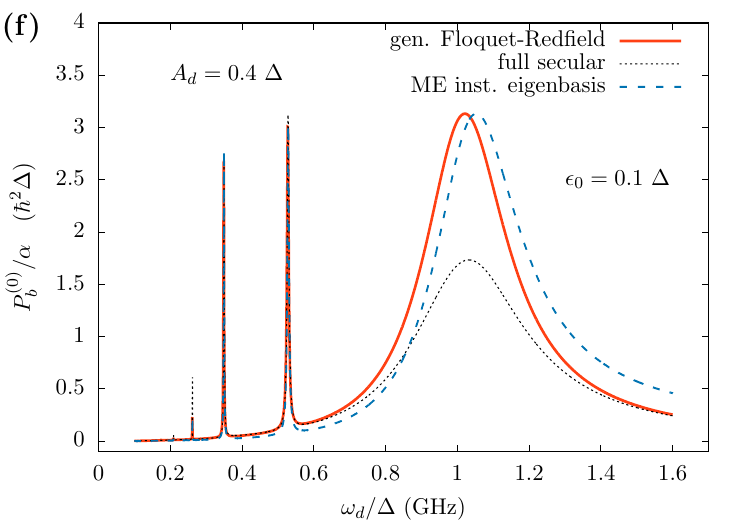}
\caption{Time-averaged, steady-state heat current $P_{\rm b}(t)$  \emph{vs.} drive frequency for two values of the drive amplitude, at zero and finite static bias. Dashed-blue lines: master equation in the instantaneous eigenbasis, Eqs.~\eqref{MEad}-\eqref{power}. The solution is propagated using a 4th-order Runge-Kutta numerical scheme up to the steady state and averaged over several drive periods. 
Comparison with the GFR approach,  Eqs.~\eqref{FBME_Fourier} and~\eqref{Pb_k0} (solid red line). The results of the full secular master equation approach, Eqs.~\eqref{ME_ss_secular}-\eqref{Pb_ss_secular} (dashed, gray lines), are also shown for reference. Other parameters are as in Fig.~\ref{simulationTLSA04}. The discrepancy at high frequency and larger $A_{\rm d}$ between Eqs.~\eqref{MEad}-\eqref{power} (blue, dashed lines) and the GFR equation is  due to  the large value of the parameter $a=\omega_{\rm d}A_{\rm d}/\Delta^2$ which signals that the drive regime is outside of the validity of the approach, see the comment below Eq.~\eqref{MEad}.} 
\label{simulationTLS_instantaneous}
\end{center}
\end{figure}

In Fig.~\ref{simulationTLS_instantaneous}, we show the results for the steady-state, time-averaged heat current to the bath from Eqs.~\eqref{MEad}-\eqref{power}. Comparison with the full Floquet-Redfield master equation shows that the approach performs very well at low frequency and/or low drive amplitude, as expected. At resonance, $\omega_{\rm d}=\bar\omega_{\rm q}$, it also captures the magnitude of the heat maximum, whereas, as shown above, see Fig.~\ref{simulationTLSA04}, the full secular master equation underestimates the maximum value of $P_{\rm b}$. On the other hand, the approach based on the master equation in the instantaneous eigenbasis requires the longest implementation time, also compared to the GFR approach, because of the required propagation up to the steady state for each value of the drive frequency. 
In Sec.~\ref{comparison_analytical_TLS} below we compare the numerical evaluation of the heat current to the bath using Eqs.~\eqref{MEad}-\eqref{power} with the one obtained analytically solving Eqs.~\eqref{MEad}.

\subsection{Fractional drive frequencies and selection rule}
\label{selection_rule}

The master equation in the instantaneous eigenbasis, Eq.~\eqref{MEad}, also provides an intuitive explanation for the multi-photon peaks around fractions of the average qubit frequency $\bar\omega_{\rm q}$, namely $\omega_{\rm d}\simeq \bar\omega_{\rm q}/l$ neglecting renormalization effects. In particular, it explains why the even fractional peaks are suppressed at zero static bias, $\epsilon_0=0$. 
For $q<1$, we write 
\begin{equation}\label{lambda(t)}
{\lambda(t)=\frac{\dot{q}}{1+q^2}=\dot{q}\sum_{k=0}^{\infty}(-q^2)^k}\;,
\end{equation}
which yields {higher} harmonics besides the basic $\omega_{\rm d}$. At the symmetry point, $\epsilon_0=0$, we have 
\be{lambda_expansion}
\lambda(t)=&\omega_{\text d}\frac{A_{\text d}}  {\Delta}\cos(\omega_{\text d} t)\sum_{k=0}^{\infty}(-1)^{k}\left( \frac{A_{\text d}}{\Delta}\right)^{{2k}} \sin^{2k}(\omega_{\text d}t)\\
=&
\omega_{\text d}\sum_{k=0}^{\infty}(-1)^{k}\left( \frac{A_{\text d}}{\Delta}\right)^{2k+1} 
\sum_{m=0}^k C_{2m+1}\cos[(2m+1)\omega_{\text d}t]\;,
\ee
which displays only odd harmonics with decreasing amplitude~\cite{Thomas2023}. Here, $C_i$ are expansion coefficients. For $\epsilon_0\neq 0$ also even harmonics contribute, see the first terms of the expansion given in Eq.~\eqref{lambda_approx} below, which produce peaks at even fractions of the qubit frequency.

\subsection{Analytical solution in the weak drive regime}\label{comparison_analytical_TLS}

In this section, we solve analytically the master equation~\eqref{MEad} in the instantaneous eigenbasis with a construction that explicitly renders the origin of the multi-photon resonances at fractional frequencies $\omega_{\rm d}=\bar\omega_{\rm q}/l$.
In the regime of weak drive ($A_{\rm d},\epsilon_0\ll \Delta$), the master equation~\eqref{MEad} becomes
\be{MEapprox}
\partial_t \rho_{00}(t)=& -\lambda(t) \mathfrak{Im}\left(e^{\ii \bar\omega_{\rm q} t} \rho_{01}(t)\right)-\Gamma_\Sigma \rho_{00}(t)+\Gamma_\downarrow\;,\\
\partial_t \rho_{01}(t)=& \; \ii\lambda(t) e^{-\ii \bar\omega_{\rm q} t} \left(\rho_{00}(t)-\frac{1}{2}\right) -\frac{1}{2}\Gamma_\Sigma \rho_{01}(t)\;,
\ee
where, expanding as in Eq.~\eqref{lambda_expansion} and defining $a=\omega_{\rm d}A_{\rm d}/\Delta^2$,
\be{lambda_approx}
\lambda(t)\simeq & a\Delta\left\{\left[1-\left(\frac{\epsilon_0}{\Delta}\right)^2 -\frac{3}{4}\left(\frac{A_{\rm d}}{\Delta}\right)^2 \right]\cos{\omega_{\rm d}t}-\frac{\epsilon_0 A_{\rm d}}{\Delta^2}\sin{2\omega_{\rm d}t}-\frac{1}{4}\left(\frac{A_{\rm d}}{\Delta}\right)^2\cos{3\omega_{\rm d}t}\right\}\\
\equiv &
\lambda_1\cos{\omega_{\rm d}t}+\lambda_2\sin{2\omega_{\rm d}t}+\lambda_3\cos{3\omega_{\rm d}t}
\ee
In Eq.~\eqref{MEapprox}, 
we made the approximation $\varphi(t)\simeq \bar\omega_{\rm q}t$, where $\bar\omega_{\rm q}$ is the one-period average of $\omega_{\rm q}(t)$ whose weak-drive limit reads $\bar\omega_{\rm q}\simeq \Delta + \frac{1}{2}\epsilon_0^2/\Delta+\frac{1}{4}A_{\rm d}^2/\Delta$. This provides a frequency renormalization induced by the drive. The rates $\Gamma_{\uparrow/\downarrow}$ in Eq.~\eqref{MEapprox} are time-independent: they are given by Eq.~\eqref{ratesTLS} upon replacing the time-dependent transition matrix elements $|Q_{01}(t)|^2$ given by Eq.~\eqref{matrix_element}, with $|\bar Q_{01}|^2=\Delta^2/\bar\omega_{\rm q}^2$ and setting $W_{10}=W(\bar \omega_{\rm q})$, see Eq.~\eqref{Re_W_omega}.

\begin{figure}[ht!]
\begin{center}
\includegraphics[width=0.44\textwidth,angle=0]{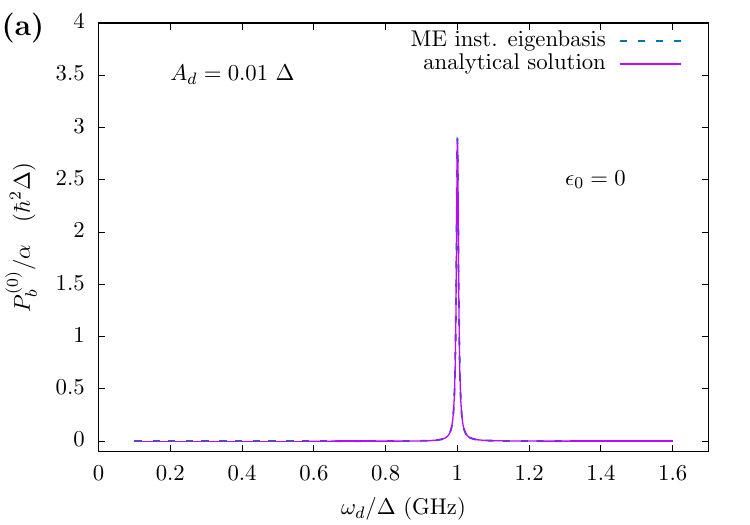}
\includegraphics[width=0.44\textwidth,angle=0]{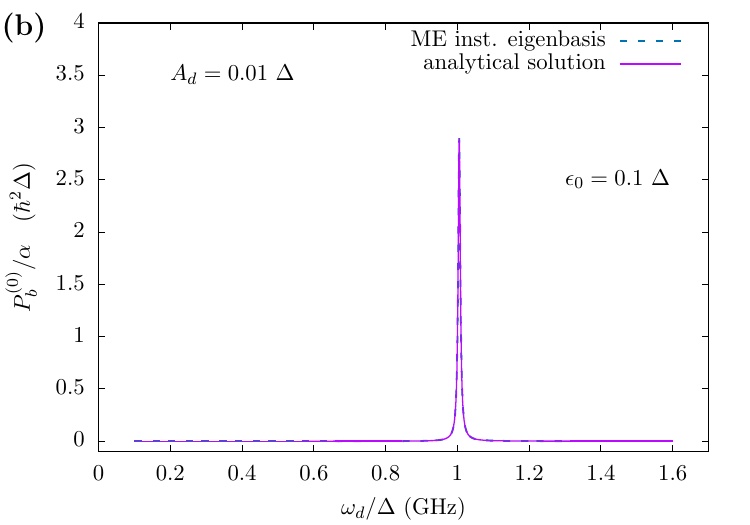}\\
\includegraphics[width=0.44\textwidth,angle=0]{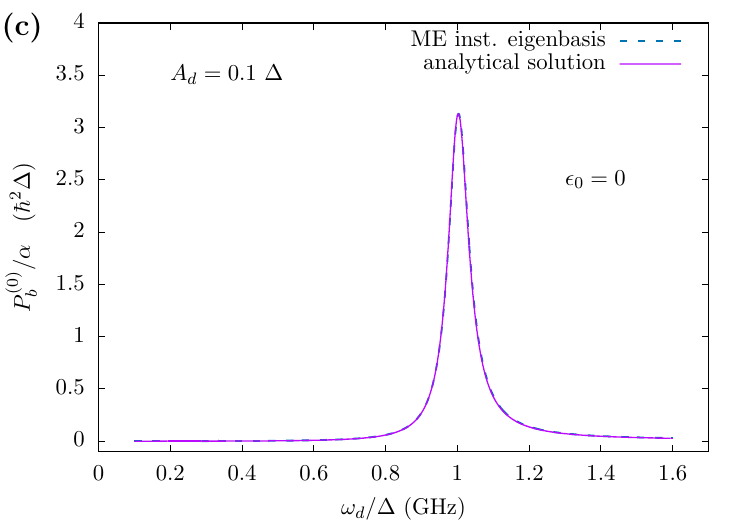}
\includegraphics[width=0.44\textwidth,angle=0]{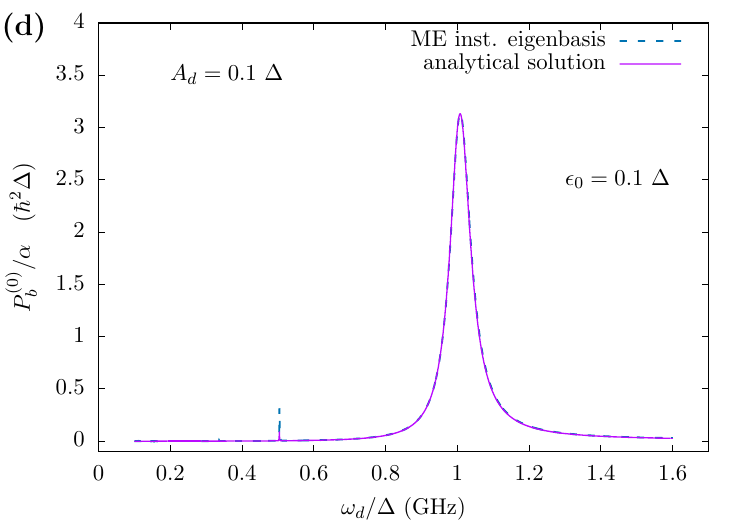}\\
\includegraphics[width=0.44\textwidth,angle=0]{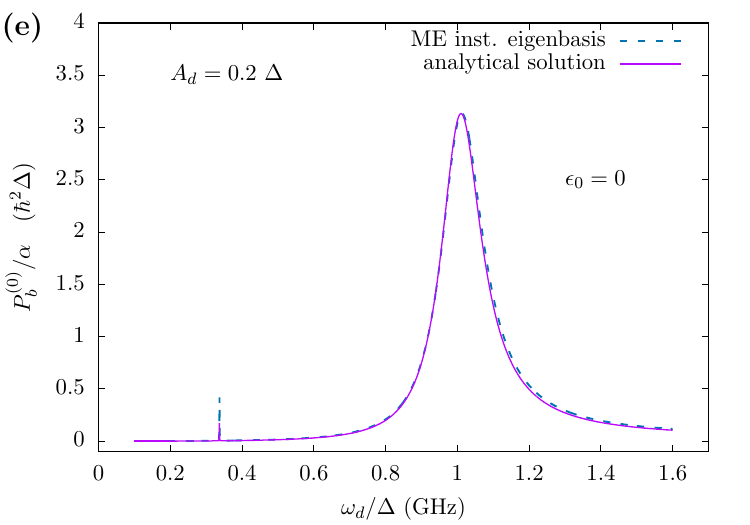}
\includegraphics[width=0.44\textwidth,angle=0]{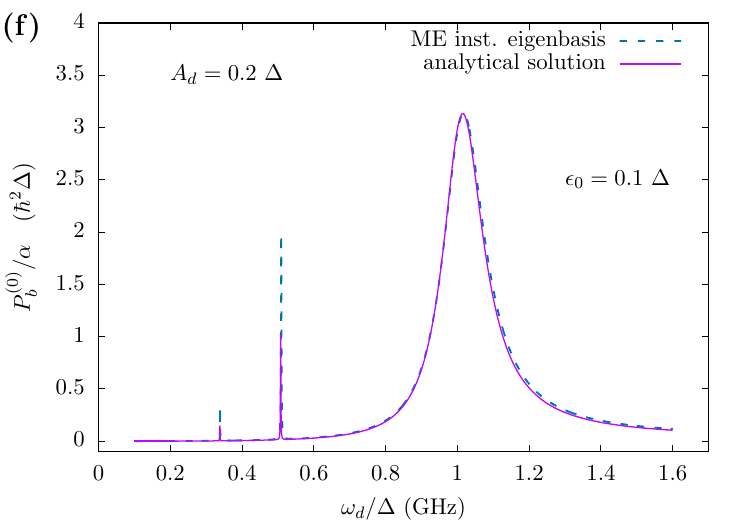}\\
\includegraphics[width=0.44\textwidth,angle=0]{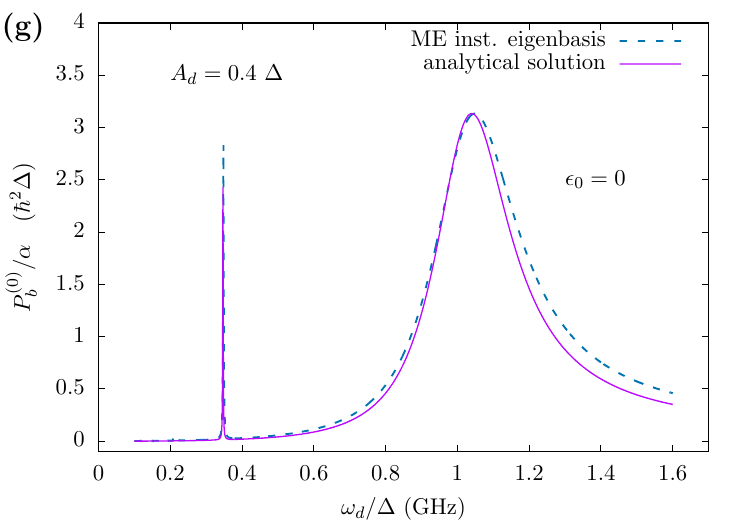}
\includegraphics[width=0.44\textwidth,angle=0]{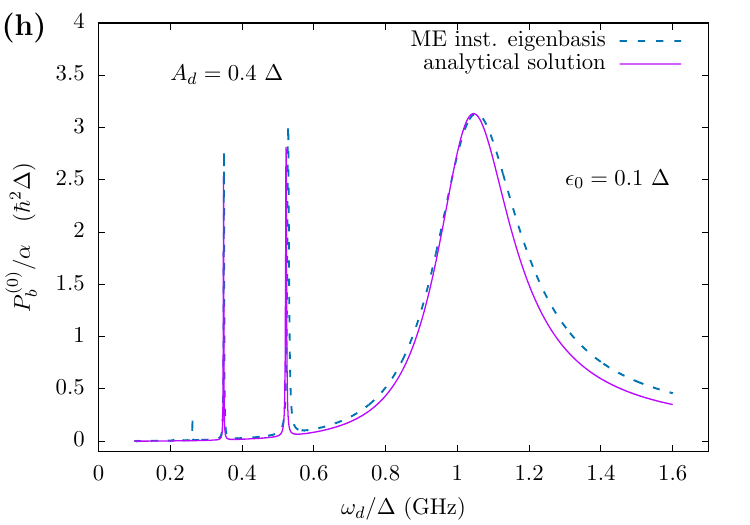}
\caption{Average steady-state heat current to the bath \emph{vs.} drive frequency at different driving amplitudes and for zero and finite static bias. Analytical calculation, Eqs.~\eqref{Approx_solution}-\eqref{Pb_analytical} (solid lines), using the weak-drive expansion for the one-period averaged qubit frequency $\bar\omega_{\rm q}$. Comparison with the numerical solution of the time-dependent master equation in the instantaneous eigenbasis (dashed lines),  Eqs.~\eqref{MEad}-\eqref{power}. Other parameters are as in Fig.~\ref{simulationTLSA04}. } 
\label{plots_analytics}
\end{center}
\end{figure}

Solving for $\rho_{01}$ with $\rho_{01}(0)=0$, assuming $\rho_{00}(t\rightarrow \infty)\simeq {\rm const.}$ and retaining only the terms which are large around resonances,  we find
\be{Approx_solution}
\rho_{01}'(t)\simeq &\left(\rho_{00}-\frac{1}{2}\right)
\left[-\lambda_1\frac{\Gamma_\Sigma\sin\omega_{1-}t}{\Gamma_\Sigma^2 +4\omega^2_{1-}}
+\lambda_2\frac{\Gamma_\Sigma\cos\omega_{2-}t}{\Gamma_\Sigma^2 +4\omega^2_{2-}}
-
\lambda_3\frac{\Gamma_\Sigma\sin\omega_{3-}t}{\Gamma_\Sigma^2 +4\omega^2_{3-}}
\right]\\
\rho_{01}''(t)\simeq &\left(\rho_{00}-\frac{1}{2}\right)
\left[
\lambda_1\frac{\Gamma_\Sigma\cos \omega_{1-}t}{\Gamma_\Sigma^2 +4\omega^2_{1-}}
+\lambda_2\frac{\Gamma_\Sigma\sin\omega_{2-}t}{\Gamma_\Sigma^2 +4\omega^2_{2-}}
+
\lambda_3\frac{\Gamma_\Sigma\cos\omega_{3-}t}{\Gamma_\Sigma^2 +4\omega^2_{3-}}
\right]\\
\rho_{00} \simeq & \frac{\Gamma_\downarrow/\Gamma_\Sigma + X/2}{1+X} \qquad\quad X=\frac{\lambda_1^2/2}{\Gamma_\Sigma^2 +4\omega^2_{1-}} + 
\frac{\lambda_2^2/2}{\Gamma_\Sigma^2 +4\omega^2_{2-}} +
\frac{\lambda_3^2/2}{\Gamma_\Sigma^2 +4\omega^2_{3-}}\;,
\ee
where $\omega_{n\pm}=n\omega_{\rm d}\pm\bar\omega_{\rm q}$. The solution for $\rho_{00}$ is obtained by substituting those for $\rho'_{01}(t)$ and $\rho''_{01}(t)$ back in the equation for $\rho_{00}(t)$, Eq.~\eqref{MEapprox}, and retaining only the constant terms: it shows that the coherences are of first order in $a$ while the first correction to the ground state population is of second order. 
The number of terms in the sum for the function $X$ reflects the number of terms retained in the expansion of $\lambda(t)$, Eq.~\eqref{lambda_approx}. In addition, this solution explicitly shows via the definitions of  $\lambda_i$, Eq.~\eqref{lambda_approx}, how the peaks at even fractions of $\bar\omega_{\rm q}$ are suppressed for $\epsilon_0=0$. \\
\indent From Eq.~\eqref{power}, the one-period average of the heat current to the bath at the steady state reads then 
\begin{equation}\label{Pb_analytical}
P_{\rm b}^{(0)}\simeq \hbar \bar\omega_{\rm q}[\Gamma_\downarrow -\Gamma_\Sigma \rho_{00}]\;.
\end{equation}
In Fig.~\ref{plots_analytics}, we plot the steady-state averaged heat current to the bath obtained according to the analytical solution, Eqs.~\eqref{Approx_solution} and~\eqref{Pb_analytical}, and compare it with the full numerical solution of the time-dependent master equation with time-dependent rates, Eqs.~\eqref{MEad}-\eqref{power}. We find that, for not-too-large drive amplitude, the analytical solution captures the behavior of the numerics, even at resonance, reproducing the fractional peaks produced by multi-photon processes at $\omega_{\rm d}=\bar\omega_{\rm q}/l$, with $l>1$, up to $l=n+1$, where $n$ is the order in $A_{\rm d}/\Delta$ retained in the expansion of $\lambda(t)$ ($n=2$ in the present treatment).\\
\indent From the analytical solution for the steady-state population $\rho_{00}$ in the instantaneous eigenbasis we can estimate height and width of the heat current peaks. At $\omega_{\rm d}=\bar\omega_{\rm q}$, for $\Gamma_\Sigma\ll A_{\rm d}$, we have $\rho_{00}\simeq 1/2$, yielding $P_{\rm b}^{(0)}\simeq \frac{\hbar \bar\omega_{\rm q}}{2}(\Gamma_\downarrow -\Gamma_\uparrow)$. At $\omega_{\rm d}=\bar\omega_{\rm q}/n$, for large $n$, such that $\Gamma_\Sigma > A_{\rm d}^n$, the peaks are suppressed because $\rho_{00}\simeq \Gamma_\downarrow/\Gamma_\Sigma$ and thus $P_{\rm b}^{(0)}\simeq 0$. Assuming a sufficiently low temperature, yielding $\Gamma_\downarrow/\Gamma_\Sigma\sim 1$, and that $\rho_{00}\simeq 1/2$, we have for the width at half maximum of the resonant peaks at $\omega_{\rm d}=\bar\omega_{\rm q}/n$, with $n=1,2,$ and $3$, 
\[
(\delta\omega_{\rm d})_1\simeq \frac{\bar\omega_{\rm q} A_{\rm d}/2}{\Delta\sqrt{2\frac{\Gamma_\downarrow}{\Gamma_\Sigma}-\frac{3}{2}}}\;,
\quad
(\delta\omega_{\rm d})_2\simeq \frac{\bar\omega_{\rm q}\epsilon_0 A_{\rm d}^2/8}{\Delta^3\sqrt{2\frac{\Gamma_\downarrow}{\Gamma_\Sigma}-\frac{3}{2}}}\;,
\qquad
(\delta\omega_{\rm d})_3\simeq \frac{\bar\omega_{\rm q} A_{\rm d}^3/72}{\Delta^3\sqrt{2\frac{\Gamma_\downarrow}{\Gamma_\Sigma}-\frac{3}{2}}}\;,
\]
respectively.
While the width of the peak at the $n$-th fractional frequency is proportional to the $n$-th power of the driving amplitude $A_{\rm d}$, the above expressions display uniform thermal broadening (the widths increase uniformly with increasing temperature) across the different multi-photon resonances. 
The approximate analytical solution, Eqs.~\eqref{Approx_solution}-\eqref{Pb_analytical}, also allows us to account for the nonlinear behavior of $P_{\rm b}^{(0)}/\alpha$ as a function of $\alpha$ shown in Fig.~\ref{simulationTLSMarkovian}.\\
Let us introduce the $\alpha$-independent quantities $\gamma_i=\Gamma_i/\alpha$ and $r =\Gamma_\downarrow/\Gamma_\Sigma$. Close to a resonance $\omega_{\rm d}\simeq \bar \omega_{\rm q}/n$, the solution we have
\be{rho_vs_alpha_analytical}
\rho_{00} \simeq & \frac{r + X/2}{1+X}\;, \qquad\quad X\simeq\frac{\lambda_n^2/2}{\alpha^2\gamma_\Sigma^2 +4\omega^2_{n-}}\simeq \frac{1}{\alpha^2} \frac{\lambda_n^2}{2\gamma_\Sigma^2}\;.
\ee
which, defining $x_n:=\lambda_n^2/2\gamma_\Sigma^2$, gives for the power to the bath
\be{Pb_vs_alpha_analytical}
\frac{P_{\rm b}^{(0)}}{\alpha}\simeq & \hbar \bar\omega_{\rm q}\gamma_\Sigma[r - \rho_{00}] \simeq\hbar \bar\omega_{\rm q}\gamma_\Sigma\frac{x_n(r-1/2)}{\alpha^2+x_n}\simeq \hbar \bar\omega_{\rm q}\gamma_\Sigma(r-1/2)(1-\alpha^2/x_n)\;.
\ee
This gives the quadratic correction in $\alpha$ of the heat current scaled with $\alpha$. 
Note that this mechanism operates within a Markovian second-order theory and does not involve higher-order corrections.

Far enough from resonance, such that $X<1$, we can expand the denominator $1+X$ in the expression for $\rho_{00}$ and take $\lambda(t)$ to lowest order in the drive, namely $\lambda_1\simeq a\Delta$, $\lambda_2=\lambda_3\simeq0$, see Eq.~\eqref{lambda_approx}. We obtain, for $A_{\rm d}\lesssim \Gamma_\Sigma$
\be{}
\rho_{00}\simeq &\;\frac{\Gamma_\downarrow}{\Gamma_\Sigma}-\frac{1}{2}\left(\frac{\omega_{\rm d}A_{\rm d}}{\Delta\Gamma_\Sigma}\right)^2\left(\frac{\Gamma_\downarrow}{\Gamma_\Sigma}-\frac{1}{2}\right)
\frac{1}{1+\left(\frac{2}{\Gamma_\Sigma}\right)^2(\omega_{\rm d}-\bar\omega_{\rm q})^2}\\
(k_B T \ll \hbar\Delta)\quad \simeq &\;1-\left(\frac{\omega_{\rm d}A_{\rm d}}{2\Delta\Gamma_\Sigma}\right)^2
\frac{1}{1+\left(\frac{2}{\Gamma_\Sigma}\right)^2(\omega_{\rm d}-\bar\omega_{\rm q})^2}\;,
\ee
which displays a single Lorentzian dip at resonance, $\omega_{\rm d}-\bar\omega_{\rm q}$, and does not capture the multi-photon resonances at fractional frequencies.

\section{Conclusions}

In this work, we provided detailed derivations of master equation approaches for calculating the heat current in driven quantum systems based on different approximations and applied them to the driven spin-boson model. We started from the weak-coupling (Born) generalized Floquet-Redfield (GFR) equation, which is non-Markovian and nonsecular. Its periodic steady state, and that of the heat current, can be rendered by a set of coupled, time-local equations in Fourier space. We also derived the Markovian approximated version of the GFR equation, the Floquet-Redfield equation, and, in turn, the full secular master equation and the respective expressions for the heat current. Applying these methods to the driven, ohmic spin-boson model, we find that the results from the Floquet-Redfield equation depart from those of the GFR equation already at weak system-bath coupling. Agreement with converged HEOM calculations supports the accuracy of the GFR predictions in the parameter regimes tested.
This shows that applying the Markovian approximation, even at weak but finite coupling, can deteriorate the predictions of the GFR. The full secular master equation provides a simple, numerically efficient method for heat transport in driven quantum systems, within the validity of the Markovian theory, but its accuracy is further limited by the drive parameters. The results from these master equations are shown to reproduce the physics of the multi-photon peaks of the heat current to the bath at low drive frequencies. We also presented the derivation of a master equation for the driven spin-boson model in the instantaneous eigenbasis, which accurately reproduce the low-frequency, multi-photon physics and remains accurate in the whole drive frequency range for not too strong drive. The analytical solution of this master equation reproduces the numerics and provides a simple description of the various features of heat transport in the driven spin-boson model, notably the peaks of heat current for driving frequencies that match integer fractions of the (average) qubit frequency and the suppression of the even fractional peaks when the driving is centered at the qubit symmetry point. Our work provides a method to study the steady-state heat transport in driven quantum systems without resorting to the usual Markovian or secular approximations, and makes the comparison with alternative approaches, clarifying the nature and impact of the approximations involved.

\section*{Acknowledgements}

The authors thank Milena Grifoni, Sigmund Kohler, and Paolo Muratore Ginanneschi for insightful discussions on the topics of open quantum systems and Floquet theory, and Katja Lindell for suggestions on how to improve the figures. The authors acknowledge the use of Aalto AI Assistant (OpenAI models, provided by Aalto University) for writing and debugging the code used in this study and revising the manuscript.  All AI-assisted work was carried out under the authors’ close supervision. This work is financially supported by the Research Council of Finland Centre of Excellence programme grant 336810 and grant 349601 (THEPOW). We acknowledge QuantERA II Programme that has received funding from the EU’s H2020 research and innovation programme under the GA No 101017733. B.K. acknowledges funding from the European Union’s Research and Innovation Programme, Horizon Europe, under the Marie Sklodowska-Curie Grant Agreement No. 101150440 (TcQTD).


\newpage

\appendix

\section{Caldeira-Leggett model}
\label{CaldeiraLeggett_model}

The Caldeira-Leggett model describes an open system bilinearly interacting via the operators $\hat X_{\rm b}$ with a set of $N_{\rm B}$ independent heat baths, each formed of $N_{\rm b}$ noninteracting harmonic oscillators, according to the total Hamiltonian~\cite{Caldeira1981,Caldeira1983,Ingold1998}
\be{H_Caldeira_Leggett}
\hat H(t)=& \hat H_{\rm S}(t)+
\hat{H}_{B}+\hat{H}_{SB}\\
=&H_{\rm S}(t)+\frac{1}{2}\sum_{\rm b}^{N_{\rm B}}\sum_{j=1}^{N_{\rm b}} \left[
\frac{\hat{p}^{2}_{{\rm b}j}}{m_{j}}+m_{{\rm b}j}\omega^{2}_{{\rm b}j}\left ( \hat{x}_{{\rm b}j}
-\frac{c_{{\rm b}j}} {m_{{\rm b}j}\omega^{2}_{{\rm b}j} } \hat{X}_{\rm b} \right )^{2}\right].
\ee
Note that the term proportional to $\hat X_{\rm b}^2$ in the system-bath interaction is introduced as a counterterm to preserve the prescribed system potential, balancing the additional static restoring-force renormalization from the bilinear coupling. This counterterm is absorbed by the system Hamiltonian in a standard partition with system, bath, and system-bath coupling terms, see below.

The spectral density function $J(\omega)$ of a bath fully characterizes the bath and its couplings to the open system. It is defined as (omitting the bath index)
\begin{equation}\label{J}
J(\omega)=\frac{\pi}{2}\sum_{j=1}^{N}\frac{c_{j}^{2}}{m_{j}\omega_{j}}\delta(\omega-\omega_{j})\;.
\end{equation}
Let us introduce the coupling constants with dimensions of angular frequency 
\[
\lambda_j=-\frac{x_0}{\sqrt{2\hbar m_j \omega_j}}c_j\;,
\]
where $x_0$ is a reference length scale. Note that since the coupling coefficients appear as $\lambda_j^2$ in the spectral density function, the minus sign in the definition is immaterial.
It is customary to define the modified spectral density function $G(\omega)$, with dimensions of frequency, as~\cite{Weiss2012} 
\be{G}
G(\omega):=\sum_{j=1}^{N}\lambda_j^2\delta(\omega-\omega_{j})=\frac{x_0^2}{\pi\hbar}J(\omega)\;.
\ee
If the system is a particle of mass $M$, in the continuum limit, the spectral density function of a so-called ohmic bath reads $J(\omega)=M\gamma\omega$, where $\gamma$ is the memoryless friction kernel that measures the overall system-bath coupling. This gives $G(\omega)=\alpha\omega$, where $\alpha=M\gamma x_0^2/\pi\hbar$ is the dimensionless system-bath coupling strength.
If the open system coupled to the bath is a harmonic oscillator of frequency $\omega_{\text r}$, we have $x_0=\sqrt{\hbar/2M\omega_{\text r}}$.\\
\indent Using $\hat{x}_{{\rm b}j}=\sqrt{\hbar/2m_{{\rm b}j}\omega_{{\rm b}j}}(\hat b_{{\rm b}j}+\hat b_{{\rm b}j}^\dag)$ and defining the dimensionless system position operator $\hat{Q}_{\rm b}$ via $\hat{X}_{\rm b}=x_{{\rm b}0}\hat{Q}_{{\rm b}}$, we can write the Hamiltonian~\eqref{H_Caldeira_Leggett} in the standard form
\be{HCL}
\hat H(t)=& \hat H_{\rm S}(t)+  \sum_{\rm b}\sum_{j=1}^{N_{\rm b}}\left[\hbar\omega_j \left(\hat b_j^\dag \hat b_j+\frac{1}{2}\right)\right]_{\rm b} -\sum_{\rm b}\hat{Q}_{\rm b}\sum_{j=1}^{N_{\rm b}} \left[x_0c_j\sqrt{\frac{\hbar}{2m_j\omega_j}}(\hat b_j+\hat b_j^\dag)\right]_{\rm b}+\sum_{\rm b}\hat A^2_{\rm b}\\
=&\tilde{H}_{\rm S}(t)+\sum_{\rm b}\hat H_{\rm b} + \sum_{\rm b}\hat{Q}_{\rm b}\sum_{j=1}^{N_{\rm b}}\hbar \left[\lambda_j(\hat b_j+\hat b_j^\dag)\right]_{\rm b}\\
=&\tilde{H}_{\rm S}(t)+\hat H_{\rm B} +\hat H_{\rm SB}\;.
\ee
Here, we have defined the renormalized system Hamiltonian $\tilde{H}_{\rm S}(t)=\hat H_{\rm S}(t)+\sum_{\rm b}\hat A^2_{\rm b}$, the bath Hamiltonian $\hat H_{\rm B}=\sum_{\rm b} \hat H_{\rm b}$, and the system-baths coupling term
\be{Hterms}
\hat H_{\rm SB}=&\;\sum_{\rm b}\hat{Q}_{\rm b}\otimes\hat{B}_{\rm b}\;,\qquad{\rm with}\qquad \hat{B}_{\rm b}:=\sum_{j=1}^{N} \hbar\left[\lambda_j(\hat b_j+\hat b_j^\dag)\right]_{\rm b}\;.
\ee
The operator that renormalizes $\hat H_{\rm S}$ is given by, see Eqs.~\eqref{H_Caldeira_Leggett}-\eqref{G}, (bath index ${\rm b}$ omitted)
\begin{equation}\label{A2}
\hat A^2=\sum_{j=1}^{N}\frac{c_{j}^{2}}{2 m_{j}\omega_{j}^2}\hat{X}^2=\frac{1}{\pi}\int_{0}^{\infty}d\omega\frac{J(\omega)}{\omega}\hat{Q}^2x_0^2 = \hbar\int_{0}^{\infty}d\omega\frac{G(\omega)}{\omega}\hat{Q}^2\equiv\mu \hat Q^2 \;.
\end{equation}
For a ohmic spectral density function with Drude cutoff at frequency $\omega_D$, i.e.  \hbox{$G(\omega)=\alpha\omega/[1+(\omega/\omega_D)^2]$}, we have $\mu=(\pi/2)\alpha\hbar\omega_D $. Similarly, an exponential cutoff function with cutoff frequency $\omega_c$ yields $\mu=\alpha\hbar\omega_c$. 
Note that if the system operator coupling to a bath is a qubit's Pauli operator, e.g. $\hat{Q}_{\rm b}=\sigma_z$, then $\hat{Q}^2_{\rm b}$ is proportional to the identity and the term $\hat A^2_{\rm b}$ merely constitutes a constant shift in the Hamiltonian.\\
\indent In the case of momentum-momentum coupling the total Hamiltonian reads~\cite{Cuccoli2001,Rastelli2016}
\be{Htot_momentum}
\hat H(t)=H_{\rm S}(t)+\frac{1}{2}\sum_{\rm b}^{N_{\rm B}}\sum_{j=1}^{N_{\rm b}} \left[
\frac{1}{m_{j}}\left(\hat{p}_{j}-\frac{c_{j}} {m_{j}\omega^{2}_{j} } \hat{P}\right)^{2}+m_{j}\omega^{2}_{j} \hat{x}_{j}^{2}
 \right]_{\rm b}\;,
\ee
and the same definitions as above apply, with the variation
\be{HSB_momentum}
\hat{Q}_{\rm b}=\ii(
\hat a^\dag-\hat a)\qquad {\rm and}\qquad\hat{B}_{\rm b}:=\ii \sum_{j=1}^{N_{\rm b}} \hbar\left[\lambda_j(\hat b_j^\dag-\hat b_j)   \right]_{\rm b}\;.
\ee
As noted in~\cite{Gelin2021}, the  construction of the equations describing the open system dynamics  in the case of position-position coupling, applies directly to the case of linear coupling of the system
to the bath momenta.
This is essentially because the bath correlation function $\langle \hat{B}_{\rm b}(t)\hat{B}_{\rm b}(0)\rangle$
is the same in the two cases, see also~\cite{Magazzu2024PRB}.
Here, the bath operators are defined as $\hat{B}:=\sum_j \hbar\lambda_j(\hat b_j + \hat b_j^\dag)$ or $\hat{B}:=\ii\sum_j \hbar\lambda_j(\hat b_j^\dag - \hat b_j)$, according to the system-bath coupling mechanism. Moreover, for independent baths, $\langle \hat{B}_{\rm b}(t)\hat{B}_{\rm b'}(0)\rangle=C_{\rm b}(t)\delta_{\rm b b'}$. The time evolution of $\hat{B}_{\rm b}(t)$ is with respect to the free bath Hamiltonian and the expectation value is calculated with respect to the thermal state $\rho_B$ of the baths. Independently of the coupling mechanism, the result is 
\be{}
C_{\rm b}(t):=&\;\langle \hat B_{\rm b}(t)\hat B_{\rm b}(0)\rangle\\
=&\;\hbar^2\sum_{j,j'} \lambda_j\lambda_{j'}\left[e^{-\ii\omega_j t}\langle \hat b_j \hat b_{j'}^\dag \rangle +e^{\ii\omega_j t}\langle \hat b_j^\dag \hat b_{j'} \rangle \right]\\
=&\;\hbar^2\sum_{j,j'} \lambda_j^2\left[e^{-\ii\omega_j t}\langle \hat b_j \hat b_{j}^\dag \rangle\delta_{jj'} +e^{\ii\omega_j t}\langle \hat b_j^\dag \hat b_{j} \rangle\delta_{jj'} \right]\\
=&\;\hbar^2\int_0^\infty d \omega\;G_{\rm b}(\omega)\left[\coth\left(\frac{\beta_{\rm b}\hbar\omega}{2}\right)\cos(\omega t)-{\rm i}\sin(\omega t)\right]\;,
\ee
where we have used the solution  $\hat b_j(t)=\hat b_j \exp(-\ii \omega_j t)$ of the Heisenberg equation in the case of noninteracting baths, the Bose-Einstein distribution $n(\omega_j)=[\exp(\beta\hbar\omega_j)-1]^{-1}$, and Eq.~\eqref{G}.
The dissipative evolution of the open system can be expressed in terms of the one-sided Fourier transform of the bath correlation function
\be{W_omega}
W_{\rm b}(\omega)=\int_0^\infty dt\; C_{\rm b}(t) e^{-\ii \omega t}\;,
\ee
whereas the power spectrum of the bath fluctuations is given by 
$
S_{\rm b}(\omega)=\int_{-\infty}^\infty dt\; C_{\rm b}(t) e^{\ii \omega t}=2{\rm Re}W_{\rm b}(-\omega)$.
This can be obtained for fairly general bath spectral density functions $G_{\rm b}(\omega)$ using the Sokhotski-Plemelj theorem, yielding
\be{Re_W_omega}
{\rm Re}W_{\rm b}(\omega)/\hbar^2=&\;\frac{\pi}{2} G_{\rm b}(|\omega|)\left[\coth\left(\frac{ \beta_{\rm b}\hbar|\omega|}{2}\right)-\frac{\omega}{|\omega|}\right]
=\;
\begin{cases}
\pi  G_{\rm b}(|\omega|)  n_{\text b}(|\omega|) \qquad \qquad \omega\geq 0\\
\pi  G_{\rm b}(|\omega|) [  n_{\text b}(|\omega|)+1]\qquad\omega<0
\end{cases}
\;.
\ee

For the specific case of an ohmic-Drude spectral density function $G(\omega)=\alpha\omega/[1+(\omega/\omega_c)^2]$, the function $W_{\rm b}(\omega)$ reads~\cite{Thingna2012,Magazzu2024PRB} (omitting the bath index ${\rm b}$)
\be{W_omega_Drude}
\frac{W(\omega)}{\hbar^2} =
\pi G(\omega)n(\omega)
-\ii\frac{\pi}{2}G(\omega)\left[\cot\left(\frac{\beta\hbar\omega_c}{2}\right)+\frac{\omega_c}{\omega}\right]+\ii\frac{2\pi\alpha\omega_c^2}{\hbar\beta}\sum_{k=1}^\infty
\frac{\nu_k \omega}{(\omega_c^2-\nu_k^2)(\omega^2+\nu_k^2)}\;,
\ee
where we used $G(-\omega)=-G(\omega)$ for the real part and where $\nu_n:=2\pi n/\beta\hbar$ are the Matsubara frequencies.

\newpage

\section{Derivation of the Nakajima-Zwanzig equation and the current formula for a driven dissipative system}
\label{NZ_derivation}

For the sake of completeness, in this appendix we provide the full  derivation, in the Schr\"odinger picture, of the Nakajima-Zwanzig (NZ) equation for the reduced density matrix of a driven, open quantum system. Let the system-plus-bath model be described by the total, time-dependent Hamiltonian defined in Appendix~\ref{CaldeiraLeggett_model}, namely
\be{}
\hat H(t)=\tilde{H}_{\rm S}(t)+\hat H_{\rm B} +\hat H_{\rm SB}\;,
\ee
and denote the total density matrix with $\varrho(t)$, whereas the system reduced density matrix is obtained by tracing over the baths' degrees of freedom $\rho(t)={\rm Tr}_{\rm B}\varrho(t)$. The total system is closed and evolves according to the Liouville-von Neumann equation 
\be{LvN}
  \frac{d}{dt}\varrho(t) = \mathcal{L}(t)\,\varrho(t)\;,
\ee
where the Liouville superoperator reads $ \mathcal{L}(t) \;\bullet := -\frac{\ii}{\hbar}[\hat H(t),\bullet]$.\\
\indent The projection operator method consists in choosing a reference state $\rho_{\rm B}$ for the baths such that $[\hat H_{\rm B},\rho_{\rm B}]=0$ and defining the two projection superoperators
\be{PQ}
  \mathcal{P}\;\bullet := \TrB (\bullet)\otimes \rho_{\rm B}\;,
  \quad {\rm and}\quad \mathcal{Q} := \id - \mathcal{P}\;.
\ee
These definitions imply $\mathcal{P}^2=\mathcal{P}$, $\mathcal{Q}^2=\mathcal{Q}$, and $\mathcal{P}\mathcal{Q}=\mathcal{Q}\mathcal{P}=0$.
We chose for the reference baths state the collection of thermal states at inverse temperatures $\beta_{\rm b}$ given by \hbox{$\rho_{\rm B}=\bigotimes_{{\rm b}=1}^{N_{\rm B}}\exp(-\beta_{\rm b}\hat H_{\rm b})/Z_{\rm b}$} with partition functions \hbox{$Z_{\rm b}={\rm Tr}\exp(-\beta_{\rm b}\hat H_{\rm b})$}. Applying the projection superoperators and their sum (which is equal to the identity superoperator) to Eq.~\eqref{LvN}, we obtain the two equations
\be{eqPrhoQrho}
  \frac{d}{dt}\,\mathcal{P}\varrho(t) &= \mathcal{P}\mathcal{L}(t)\mathcal{P}\varrho(t) + \mathcal{P}\mathcal{L}(t)\mathcal{Q}\varrho(t)\;,\\
  \frac{d}{dt}\,\mathcal{Q}\varrho(t) &= \mathcal{Q}\mathcal{L}(t)\mathcal{P}\varrho(t) + \mathcal{Q}\mathcal{L}(t)\mathcal{Q}\varrho(t)\;.
\ee
At this point, we define the irreducible propagator superoperator  $\mathcal{G}_{\mathcal{Q}}(t,t')$ as the solution of
\be{GQ}
  \partial_t \mathcal{G}_{\mathcal{Q}}(t,t') = \mathcal{Q}\mathcal{L}(t) \mathcal{G}_{\mathcal{Q}}(t,t'),\quad {\rm with}\quad \mathcal{G}_{\mathcal{Q}}(t,t)=\mathcal{Q}\;,
\ee
which formally reads 
\be{GQsolution}
  \mathcal{G}_{\mathcal{Q}}(t,t') = \mathcal{T}\exp\left[\int_{t'}^{t} d\tau\, \mathcal{Q}\mathcal{L}(\tau)\right]\mathcal{Q}\;,
\ee
where $\mathcal{T}$ denotes the time ordering operator. One can check that the ansatz 
\be{solutionQrho}
\mathcal{Q}\varrho(t) = \mathcal G_{\mathcal{Q}}(t,t_0)\,\mathcal{Q}\varrho(t_0)
+ \int_{t_0}^{t} dt'\, \mathcal G_{\mathcal{Q}}(t,t')\,\mathcal{Q}\mathcal{L}(t')\,\mathcal{P}\varrho(t') \;,
\ee
with $\mathcal G_{\mathcal{Q}}(t,t')$ given by Eq.~\eqref{GQsolution}, solves the second of Eqs.~\eqref{eqPrhoQrho}. Substituting this solution in the first of Eqs.~\eqref{eqPrhoQrho} yields
\be{solutionPrho}
\frac{d}{dt}\,\mathcal{P}\varrho(t) &= \mathcal{P}\mathcal{L}(t)\mathcal{P}\varrho(t) + \mathcal{P}\mathcal{L}(t)\mathcal G_{\mathcal{Q}}(t,t_0)\,\mathcal{Q}\varrho(t_0)
+ \int_{t_0}^{t} dt'\, \mathcal{P}\mathcal{L}(t)\mathcal G_{\mathcal{Q}}(t,t')\,\mathcal{Q}\mathcal{L}(t')\,\mathcal{P}\varrho(t')\;.
\ee
Tracing over the baths degrees of freedom, we obtain the general, inhomogeneous form of the NZ equation
\be{NZ_equation_general}
  \dot{\rho}(t) = \mathcal{L}_{\rm S}^{\mathrm{eff}}(t)\,\rho(t)
  + \int_{t_0}^{t} dt'\, \mathcal{K}(t,t')\,\rho(t') + \mathcal{I}(t)\;,
\ee
with the following definitions
\be{definitions}
  \mathcal{L}_{\rm S}^{\mathrm{eff}}(t) \rho(t) &\equiv \Tr_{\rm B}\left\{\mathcal{L}(t)\rho(t)\otimes\rho_{\rm B}\right\}
  = -\frac{\ii}{\hbar}\,[\tilde H_{\rm S}(t) + \hat H'_{\rm S}, \rho(t)]\;, \\
 \hat H'_{\rm S}&\equiv \TrB\big(\hat H_{\rm SB}\rho_{\rm B}\big)\;, \\
  \mathcal{K}(t,t') \rho(t') &\equiv \TrB\left\{\mathcal{L}(t)\mathcal G_{\mathcal{Q}}(t,t') \mathcal{L}(t')\rho(t')\otimes \rho_{\rm B}\right\}\;, \\
  \mathcal{I}(t) &\equiv \TrB\left\{\mathcal{L}(t)\mathcal G_{\mathcal{Q}}(t,t_0) \mathcal{Q}\varrho(t_0)\right\}\;.
\ee
In the third of Eqs.~\eqref{definitions} we have used $  \mathcal{Q}\,\mathcal G_{\mathcal{Q}}(t,t') \;=\; \mathcal G_{\mathcal{Q}}(t,t') \;=\;\mathcal G_{\mathcal{Q}}(t,t')\,\mathcal{Q}$, see Eq.~\eqref{GQsolution}.\\
\indent The NZ equation~\eqref{NZ_equation_general} is exact and derived in a rather general setting. Assuming now that the initial condition is factorized, implying $\mathcal Q\varrho(t_0)=0$ (and thus $\mathcal{I}(t)=0$) and that, without loss of generality, the interaction Hamiltonian is such that $\hat H'_{\rm S}=0$, which gives $\mathcal{L}_{\rm S}^{\mathrm{eff}}(t)=\mathcal{L}_{\rm S}(t)$, we recover, for $t_0=0$, Eq.~\eqref{NZ_equation}. Note that the kernel in  Eq.~\eqref{NZ_equation} is equivalently written as
\be{kernel_rho}
\mathcal{K}(t,t') \rho(t') &\equiv \TrB\left\{\mathcal{L}_{\rm SB}\mathcal G_{\mathcal{Q}}(t,t') \mathcal{L}_{\rm SB}\rho(t')\otimes \rho_{\rm B}\right\},
\ee
where we used the identities $\mathcal{P}\mathcal{L}_{\rm S}(t)\mathcal{Q}=0,\quad \mathcal{Q}\mathcal{L}_{\rm S}(t)\mathcal{P}=0,\quad \mathcal{P}\mathcal{L}_{\rm B}=0,\quad \mathcal{Q}\mathcal{L}_{\rm B}\mathcal{P}=0$, which follow from $[\hat H_{\rm B},\rho_{\rm B}]=0$.

\subsection*{Heat current}
\label{heat_current_derivation}

The heat current is defined as $P_{\rm b}(t)={\rm Tr}[\hat I_{\rm b}\varrho(t)]={\rm Tr}[\hat I_{\rm b}(\mathcal P + \mathcal Q)\varrho(t)]$. For the current operators relevant to this work, see Sec.~\ref{heat_current}, ${\rm Tr}_{\rm B}[\hat I_{\rm b}\mathcal P\varrho(t)]=0$. As a result $P_{\rm b}(t)={\rm Tr}[\hat I_{\rm b}\mathcal Q\varrho(t)]$. Using Eq.~\eqref{solutionQrho} for a factorized initial condition ($\mathcal Q\varrho(t_0)=0$) we obtain
\be{}
P_{\rm b}(t)={\rm Tr}_{\rm S}\int_0^t dt' \; \TrB\{\hat I_{\text b} \mathcal{G}_{\mathcal{Q}}(t,t')\mathcal{L}(t')\rho(t')\otimes\rho_B\}\;,
\ee
which coincides with Eq.~\eqref{Pb_exact}, namely $\mathcal{K}_{I,{\rm b}}(t,t')\rho(t')=\TrB \{\hat I_{\text b} \mathcal{G}_{\mathcal{Q}}(t,t')\mathcal{L}_{\rm SB}\rho(t')\otimes\rho_B\}$ upon using again, as in Eq.~\eqref{kernel_rho}, the action of the projection operators on $\mathcal{L}_{\rm S}(t)$ and $\mathcal{L}_{\rm B}$.

\section{Markovian approximation: Floquet-Redfield master equation}
\label{Markovian}

The interaction picture with respect to the free system Hamiltonian is given by $\rho_{\rm I}(t)=U_{\rm S}^\dag(t,0)\rho(t)U_{\rm S}(t,0)$.
In this picture, the Nakajima-Zwanzig equation~\eqref{NZ_equation} in the weak-coupling approximation is transformed into
\be{NZ_equation_int}
\dot{\rho_{\rm I}}(t) &= \int_0^t d\tau   U^\dag_S(t,0)\mathcal{K}(t,t-\tau)[\rho(t-\tau)]U_S(t,0)\\
&=
-\frac{1}{\hbar^2}\sum_{\rm b}\int_0^t d\tau  \Big\{\left[\hat{Q}_{\rm b,I}(t),\hat{Q}_{\rm b,I}(t-\tau)\rho_{\rm I}(t-\tau)\right]C_{\rm b}(\tau) 
+
\left[\rho_{\rm I}(t-\tau)\hat{Q}_{\rm b,I}(t-\tau), \hat{Q}_{\rm b,I}(t)\right]C_{\rm b}^*(\tau)\Big\}\;,
\ee
where $\hat{Q}_{\rm I}(t)=U^\dag_{\rm S}(t,0)\hat Q U_{\rm S}(t,0)$. 
For the heat current in terms of quantities expressed in the interaction picture we obtain
\be{A_KI_second}
P_{\rm b}(t)
&=-\frac{\ii}{\hbar}{\rm Tr}_{\rm S}\int_0^t d\tau  \Big[\hat{Q}_{\rm b,I}(t) \hat{Q}_{\rm b,I}(t-\tau) \rho_{\rm I}(t-\tau)K_{\rm b}(\tau)-\hat{Q}_{\rm b,I}(t) \rho_{\rm I}(t-\tau)\hat{Q}_{\rm b,I}(t-\tau)K_{\rm b}^*(\tau)\Big]\;.
\ee

Assuming that the interaction-picture RDM evolves much slower than the decay times $\tau_{\rm b}$ of the bath correlation functions $C_{\rm b}(t)$, we can make the substitutions $\rho_{\rm I}(t-\tau)\rightarrow \rho_{\rm I}(t)$ and, for long times ($t\gg \tau_{\rm b}$), $\int_0^t d\tau\rightarrow \int_0^\infty d\tau$, yielding
\be{BMME}
\dot{\rho_{\rm I}}(t) \simeq -\frac{1}{\hbar^2}\int_0^\infty d\tau
\sum_{\rm b}\Big\{\left[\hat{Q}_{\rm b,I}(t),\hat{Q}_{\rm b,I}(t-\tau)\rho_{\rm I}(t)\right]C_{\rm b}(\tau) 
+
\left[\rho_{\rm I}(t)\hat{Q}_{\rm b,I}(t-\tau), \hat{Q}_{\rm b,I}(t)\right]C_{\rm b}^*(\tau)\Big\}\;.
\ee
Similarly, if $t\gg \tau_{\rm b}$, for the heat current we obtain

\be{MHC}
P_{\rm b}(t)
&\simeq-\frac{\ii}{\hbar}{\rm Tr}_{\rm S}\int_0^\infty d\tau  \Big\{\hat{Q}_{\rm b,I}(t) \hat{Q}_{\rm b,I}(t-\tau) \rho_{\rm I}(t)K_{\rm b}(\tau)-\hat{Q}_{\rm b,I}(t) \rho_{\rm I}(t)\hat{Q}_{\rm b,I}(t-\tau)K_{\rm b}^*(\tau)\Big\}\;.
\ee
Projecting these expression in the basis $\{\ket{\Psi_n(0)}\}=\{\ket{u_n(0)}\}$ and using the action of the Floquet propagator given by Eq.~\eqref{floquet_propagator} we have $\bra{\Psi_n(0)}\rho_{\rm I}(t)\ket{\Psi_m(0)}=\bra{\Psi_n(t)}\rho(t)\ket{\Psi_m(t)}=\rho_{nm}(t)$ 
\be{A_FBMME}
\dot\rho_{nm}(t) \simeq
& -\frac{1}{\hbar^2}\sum_{{\rm b},n'm'}\int_0^\infty d\tau
\Big\{\sum_p\Big[\delta_{mm'}Q_{np}(t) Q_{pn'}(t-\tau)C(\tau) + \delta_{nn'}Q_{m'p}(t-\tau)Q_{pm}(t)C^*(\tau)\Big]\\ 
&\qquad \qquad\quad -
\Big[Q_{m'm}(t)Q_{nn'}(t-\tau)C(\tau) + Q_{m'm}(t-\tau)Q_{nn'}(t)C^*(\tau)\Big]\Big\}_{\rm b}\rho_{n'm'}(t)\\
= - \sum_{{\rm b},n'm'}&\frac{e^{\ii(\omega_{nm}-\omega_{n'm'})t}}{\hbar^2}\int_0^\infty d\tau
\Big\{\sum_p\Big[\delta_{mm'}\tilde Q_{np}(t) \tilde Q_{pn'}(t-\tau)e^{-\ii\omega_{pn'}\tau}C(\tau) + \delta_{nn'}\tilde Q_{m'p}(t-\tau)\tilde Q_{pm}(t)e^{\ii\omega_{pm'}\tau}C^*(\tau)\Big]\\ 
&\qquad \qquad -
\Big[\tilde Q_{m'm}(t)\tilde Q_{nn'}(t-\tau)e^{-\ii\omega_{nn'}\tau}C(\tau) + \tilde Q_{m'm}(t-\tau)\tilde Q_{nn'}(t)e^{\ii\omega_{mm'}\tau}C^*(\tau)   \Big]\Big\}_{\rm b}\rho_{n'm'}(t)
\;,
\ee
At this point, we Fourier-expand $\tilde Q(t)=\sum_{l}e^{\ii l {\omega_{\text d}} t}Q_{n m  }^{l}$ and integrate over $\tau$ or yielding
\be{FRint}
\dot\rho_{nm}(t) \simeq
 - \sum_{n'm'll'}\frac{e^{\ii(\omega_{nm}^l-\omega_{n'm'}^{l'})t}}{\hbar^2}\sum_{\rm b}
\Big\{\sum_p\Big[&\delta_{mm'} Q_{np}^{-l'}  Q_{pn'}^{l}W_{pn'}^{l} + \delta_{nn'} Q_{m'p}^{-l'} Q_{pm}^{l}[W_{pm'}^{l'}]^*\Big]\\ 
- & Q_{m'm}^{-l'} Q_{nn'}^l \Big[W_{nn'}^l + [W_{mm'}^{l'}]^*\Big]\Big\}_{\rm b}\rho_{n'm'}(t)
\;,
\ee
Substituting $\rho_{nm}(t)=e^{\ii\omega_{nm}t}\tilde\rho_{nm}(t)$ in Eq.~\eqref{FRint},  we obtain, in the basis of Floquet-modes, the Markovian Floquet-Redfield master equation
\be{}
\dot{\tilde\rho}_{nm}(t) \simeq
 -\ii\omega_{nm}\tilde\rho_{nm}(t) 
  - \sum_{n'm'll'}\frac{e^{\ii(l-l')\omega_{\rm d}t}}{\hbar^2}
\sum_{\rm b}\Big\{\sum_p\Big[&\delta_{mm'} Q_{np}^{-l'}  Q_{pn'}^{l}W_{pn'}^{l} + \delta_{nn'} Q_{m'p}^{-l'} Q_{pm}^{l}[W_{pm'}^{l'}]^*\Big]\\ 
 - & Q_{m'm}^{-l'} Q_{nn'}^l \Big[W_{nn'}^l + [W_{mm'}^{l'}]^*\Big]\Big\}_{\rm b}\tilde\rho_{n'm'}(t)
\ee
At the steady state, Fourier-expanding the RDM in the basis of Floquet modes we obtain an expression similar to Eq.~\eqref{FBME_Fourier}, namely  
\be{A_FBME_FourierMarkovian}
0^{(k)}
=-\ii\omega_{nm}^k\tilde\rho^{(k)}_{nm}+\sum_{n'm',k'}R_{nmn'm'}^{k,k'(2)M}\tilde\rho^{(k')}_{n'm'}\;,
\ee
where $M$ stands for \emph{Markovian}. Here, the nonsecular Floquet-Redfield tensor reads, cf. Eq.~\eqref{FRTensor},
\be{A_FRTensorMarkovian}
R_{nmn'm'}^{k,k'(2)M}=-\frac{1}{\hbar^2}\sum_{\rm b}\sum_{l}\Bigg\{\sum_{p }\Big[&\delta_{mm'}Q_{n p }^{-l+k-k'}Q^{l}_{p n'}W_{p n'}^{l}+ 
\delta_{nn'} Q_{m'p }^{-l+k-k'}Q^{l}_{p m}[W_{p m'}^{l-k+k'}]^*\Big]\\
 - & Q_{m'm}^{-l+k-k'}Q_{nn'}^{l}\left[W_{nn'}^{l}
+ [W_{mm'}^{l-k+k'} ]^*\right]\Bigg\}_{\rm b}\;.
\ee
The corresponding expression for the Fourier components of the heat current to bath $\rm b$ in the Markovian approximation is
\be{}
P_{\rm b}^{(k)M}=&-\frac{1}{\hbar}\sum_{n ,n ',m'}\sum_{l,k'} \Big\{
Q_{{\rm b},m'n }^{-l+k-k'}Q_{{\rm b},n n '}^{l}
\omega_{nn'}^{l} W^{l}_{{\rm b},nn'}
\tilde\rho_{n 'm'}^{(k')}
+
[Q_{{\rm b},m'n }^{-l-k-k'}Q_{{\rm b},n n '}^{l}
\omega_{nn'}^{l} W^{l}_{{\rm b},nn'}
\tilde\rho_{n 'm'}^{(k')}]^*\Big\}\;,
\ee
cf. Eq.~\eqref{Pb_k}, where the RDM elements $\tilde\rho_{n 'm'}^{(k')}$ are given by Eq.~\eqref{A_FBME_FourierMarkovian}.

\subsection{Secular approximation: Floquet-Lindblad equation}

For completeness, we show here the derivation of the Floquet-Lindblad equation, starting from the Markovian, Floquet-Redfield equation in the interaction picture. To this end, we express Eq.~\eqref{FRint}, in operatorial form ($\Omega\equiv \omega_{nm}^l$)
\be{}
\dot{\rho}_{\rm I}(t)
=
-\frac{1}{\hbar^2}\sum_{{\rm b},\Omega,\Omega'}
e^{+\ii(\Omega-\Omega')t}
\Bigl\{&
\hat Q_{{\rm b},\Omega'}^\dagger \hat Q_{{\rm b},\Omega}\rho_{\rm I}(t)W_{\rm b}(\Omega) + \rho_{\rm I}(t) \hat Q_{{\rm b},\Omega'}^\dagger \hat Q_{{\rm b},\Omega}W_{\rm b}^*(\Omega')\\
-&
\hat Q_{{\rm b},\Omega}\rho_{\rm I}(t) \hat Q_{{\rm b},\Omega'}^\dagger
\left[W_{\rm b}(\Omega)+W_{\rm b}^*(\Omega')\right]
\Bigr\}\;,
\ee
where $\hat Q_{{\rm b},\Omega}:=\sum_{n,m,l:\,\omega_{nm}^{l}=\Omega} Q_{{\rm b},nm}^{\,l}
\ket{\Psi_n(0)}\bra{\Psi_m(0)}$. Retaining in the sum only the terms with $\Omega=\Omega'$ (namely $\omega_{nm}^l=\omega_{n'm'}^{l'}$) we obtain the secular Floquet-Lindblad master equation in the interaction picture
\be{FL_int}
\dot{\rho}_{\rm I}(t)
=
-\frac{\ii}{\hbar}
[\hat H_{\rm LS},\rho_{\rm I}(t)]
+
\sum_{{\rm b},\Omega}
\gamma_{\rm b}(\Omega)
\left[
\hat Q_{{\rm b},\Omega}\rho_{\rm I}(t)\hat Q_{{\rm b},\Omega}^\dagger
-\frac{1}{2}
\left\{
\hat Q_{{\rm b},\Omega}^\dagger \hat Q_{{\rm b},\Omega},
\rho_{\rm I}(t)
\right\}
\right],
\ee
where the Lamb-shift Hamiltonian reads
\be{HLS}
\hat H_{\rm LS}
=
\sum_{{\rm b},\Omega}
s_{\rm b}(\Omega)
\hat Q_{{\rm b},\Omega}^\dagger \hat Q_{{\rm b},\Omega}\;,
\ee
and where $\gamma_{\rm b}(\Omega) :=
\frac{2}{\hbar^2}{\rm Re} W_{\rm b}(\Omega)$ and $
s_{\rm b}(\Omega) := \frac{1}{\hbar}{\rm Im} W_{\rm b}(\Omega)$. 
With reference to Eq.~\eqref{FRint}, the above equation reads, in the basis of Floquet states,
\be{FL_Floquet}
\dot\rho_{nm}(t) =
 - \frac{1}{\hbar^2} \sum_{n'm'll':\,\omega_{nm}^{l}=\omega_{n'm'}^{l'}}\sum_{\rm b}
\Big\{\sum_p\Big[&\delta_{mm'} Q_{np}^{-l'}  Q_{pn'}^{l}W_{pn'}^{l} + \delta_{nn'} Q_{m'p}^{-l'} Q_{pm}^{l}[W_{pm'}^{l'}]^*\Big]\\ 
- & Q_{m'm}^{-l'} Q_{nn'}^l \Big[W_{nn'}^l + [W_{mm'}^{l'}]^*\Big]\Big\}_{\rm b}\rho_{n'm'}(t)\;.
\ee
Note that the secular condition does not imply $\omega_{nm}=\omega_{n'm'}$ and $l=l'$ separately, unless these  extended frequencies are nondegenerate. In this case the secular approximation coincides with the more restrictive ultra-secular approximation which results, at the steady state, in Eq.~\eqref{ME_ss_secular}.

\subsection{Fast-decaying bath/slow drive regime}

Starting from Eq.~\eqref{A_FBMME} and making the further approximation $\tilde Q(t-\tau)\simeq \tilde Q(t)$, under the assumption (stronger than the Markovian approximation) that the bath correlation functions decay faster than the driving cycle, leads to the following Markovian master equation with time-dependent Floquet-Redfield tensor in the basis of Floquet modes
\be{A_FRME}
\dot{\tilde\rho}_{nm}(t) \simeq
 -\ii\omega_{nm}\tilde\rho_{nm}(t) -\frac{1}{\hbar^2}
\sum_{n'm'}&\sum_{\rm b}\Big\{\sum_p\Big[\delta_{mm'}\tilde Q_{np}(t) \tilde Q_{pn'}(t)W_{pn'} + \delta_{nn'}\tilde Q_{m'p}(t)\tilde Q_{pm}(t)W_{pm'}^*\Big]\\ 
&\qquad \qquad -
\tilde Q_{m'm}(t)\tilde Q_{nn'}(t)\Big[W_{nn'} + W_{mm'}^*\Big]\Big\}_{\rm b}\tilde\rho_{n'm'}(t)\;.
\ee
Since we are assuming here that $C(\tau)$ decays fast on the time scale of the drive, Eq.~\eqref{A_FRME} is best suited for the regime of  slow driving.
Correspondingly, under the same assumptions, Eq.~\eqref{MHC} gives for the heat current to the bath
\be{FMHC}
P_{\rm b}(t)\simeq 
&-\frac{1}{\hbar}\sum_{n ,n',m'}2{\rm Im}\Big\{ 
\tilde Q_{{\rm b}, m'n }(t)\tilde Q_{{\rm b}, n n '}(t)e^{\ii\omega_{m'n'}t}\rho_{n 'm'}(t)\int_0^\infty d\tau\; e^{-\ii\omega_{nn'}\tau}\frac{d}{d\tau}C_{\rm b}(\tau)
\Big\}\\
=&-\sum_{n ,n',m'}\hbar \omega_{nn'}2{\rm Re}\Big\{ 
\tilde Q_{{\rm b}, m'n }(t)\tilde Q_{{\rm b}, n n '}(t)\tilde\rho_{n 'm'}(t)W_{{\rm b}, nn'}/\hbar^2\Big\}\;.
\ee

\section{Numerical implementation of the Floquet eigensystem}
\label{numerical_implementation}

From the Floquet theorem, the propagation of the Floquet modes can be rendered as
\be{FModes}
\ket{\Psi_n(t)}&=e^{-\ii\varepsilon_n t}\ket{u_n(t)}=\mathcal{U}(t,0)\ket{\Psi_n(0)}=\mathcal{U}(t,0)\ket{u_n(0)}\\
\implies\qquad 
\ket{u_n(t)}&=e^{\ii\varepsilon_n t}\mathcal{U}(t,0)\ket{u_n(0)}\;.
\ee
This propagation requires the knowledge of the Floquet state propagator $\mathcal{U}(t,0)$, the quasienergies, and the Floquet modes (Floquet states) at $t=0$. This eigensystem is found via
\be{}
\ket{\Psi_n(T)}&=e^{-\ii\varepsilon_n T}\ket{u_n(T)}=e^{-\ii\varepsilon_n T}\ket{u_n(0)}=\mathcal{U}(T,0)\ket{u_n(0)}\;
\ee
which means that $\ket{u_n(0)}$ are eigenstates of the one-period propagator with eigenvalues $\lambda_n=e^{-\ii\varepsilon_n T}$, and thus $\varepsilon_n=-\arg(\lambda_n)/T$.

The propagator can in turn be obtained approximately via Trotterization of the time-evolution operator~\cite{Viebahn2020} 
\be{}
\mathcal{U}(T,0)=\mathcal{T} e^{-\frac{\ii}{\hbar}\int_0^T d t' H(t')}\simeq e^{-\frac{\ii}{\hbar}\sum_{i=0}^{N-1} H(t_i)\Delta t} = \prod_{i=0}^{N-1} e^{-\frac{\ii}{\hbar} H(t_i)\Delta t} +\mathcal{O}(\Delta t ^2)\;.
\ee
Note that the propagator at intermediate times ($t<T$) has the same structure and can be used to propagate the Floquet modes according to Eq~\eqref{FModes}.
Propagation at larger times $t=nT+s$, with $s\in [0,T)$, is obtained from the knowledge of the Floquet modes within one period: 
\be{}
\ket{\Psi_n(t)}=e^{-\ii\varepsilon_n t}\ket{u_n(t)}=e^{-\ii\varepsilon_n t}\ket{u_n(s)}\;.
\ee

\end{document}